\documentclass[twocolumn,superscriptaddress,amsmath,amssymb,aps,longbibliography,prx]{revtex4-2}
\usepackage[english,american]{babel}
\usepackage[colorlinks=true,urlcolor=blue,citecolor=blue,linkcolor=blue]{hyperref}
\usepackage{txfonts}
\usepackage{graphicx}
\usepackage{dsfont}
\usepackage{bbm}
\usepackage{booktabs}
\usepackage{amsmath}
\usepackage{algorithm}
\usepackage{algorithmicx}
\usepackage{algpseudocode} 
\usepackage{graphicx}% Include figure files
\usepackage{dcolumn}% Align table columns on decimal point
\usepackage{rotating}
\usepackage{longtable, booktabs, threeparttablex}

\usepackage{multirow}
\usepackage{threeparttable}
\usepackage{color}
\usepackage{xcolor} % Include the package
\definecolor{lightgreen}{RGB}{144,238,144} % RGB values for light green
\definecolor{lightgray}{gray}{0.75}

\renewcommand{\vec}[1]{\boldsymbol{#1}}
\newcommand{\Eq}[1]{Eq.~(\ref{#1})}
\newcommand{\Fig}[1]{Fig.~\ref{#1}}

\newcommand{\ra}[1]{\renewcommand{\arraystretch}{#1}}

\graphicspath{{figure/}}

\begin{document}

\title{Neural Canonical Transformation for the Spectra of Fluxional Molecule CH$_5^+$}

\author{Ruisi Wang}
\affiliation{Beijing National Laboratory for Condensed Matter Physics and Institute of Physics, \\Chinese Academy of Sciences, Beijing 100190, China}
\affiliation{School of Physical Sciences, University of Chinese Academy of Sciences, Beijing 100190, China}

\author{Qi Zhang}
\email{zhangqi94@iphy.ac.cn}
\affiliation{Beijing National Laboratory for Condensed Matter Physics and Institute of Physics, \\Chinese Academy of Sciences, Beijing 100190, China}

\author{Lei Wang}
\email{wanglei@iphy.ac.cn}
\affiliation{Beijing National Laboratory for Condensed Matter Physics and Institute of Physics, \\Chinese Academy of Sciences, Beijing 100190, China}

\date{\today}

\begin{abstract}
Protonated methane, $\text{CH}_5^+$, is a highly fluxional molecule with large spatial motions of the hydrogen atoms. The molecule's anharmonic effects and the delocalized wavefunction of the hydrogen atoms significantly affect the excitation spectrum of the molecule.
The neural canonical transformation (NCT) approach, which we previously developed to solve the vibrational spectra of molecules and solids, 
is a powerful method that effectively treats nuclear quantum effects and anharmonicities.
Using NCT with wavefunctions in atomic coordinates rather than normal coordinates, we successfully calculate the ground and excited states of $\text{CH}_5^+$.
We found that the wavefunctions for the ground state, as well as for low- and high-energy excited states, show preferences for the three stationary points on the potential energy surface.
This work extends the applicability of the NCT approach for calculating excited states to fluxional molecules without fixed geometry. 
\end{abstract}

\maketitle

\section{Introduction}
\label{sec:intro}

Protonated methane, CH$_5^+$, is known for its highly fluxional nature~\cite{schreiner_ch5_1993,marx_structural_1995,marx_ch_1999,asvany_understanding_2005}. 
It represents the prototypic superacid that plays a pivotal role as a reactive intermediate in acid-catalyzed electrophilic reactions~\cite{olah1995electrophilic,olah1997kekule}, and is also considered a crucial intermediate in the formation of polyatomic organic compounds within cold interstellar clouds~\cite{he03000u,talbi1992quantum}.
Although initially observed by mass spectroscopy in the early 1950s~\cite{tal1952and}, the first infrared (IR) spectrum of $\text{CH}_5^+$ was not successfully measured until 1999~\cite{white_ch_1999}, yet the observed spectral lines could not be assigned to specific rovibrational motions such as C--H stretching, H$_2$ internal rotation, or bending modes.
Another breakthrough occurred in 2015 with the measurement of experimental combination differences by Asvany et al.~\cite{asvany_experimental_2015} (with an improvement reported in Ref.~\cite{brackertz_searching_2017}), serving as another key experiment in advancing the understanding of CH$_5^+$~\cite{oka_taming_2015}.
The combination difference is the difference of two transition wavenumbers that share a common energy level. Subtracting them cancels the shared level and retrieves the energy spacing of the other two levels.

Concurrent with experimental progress, significant advancements in numerical methods have emerged. An ab initio calculation of the millimeter-wave spectrum of CH$_5^+$~\cite{bunker_theoretical_2004} and an MP2/cc-pVTZ-based potential energy surface (PES)~\cite{brown_classical_2003} were reported in 2003. 
In 2004, diffusion Monte Carlo (DMC) techniques were applied to the ground-state calculation of CH$_5^+$ with an ab initio-based PES~\cite{mccoy_ab_2004,brown_quantum_2004,thompson_ch_2005}.
Then in 2006, a full-dimensional PES based on CCSD(T)/aug-cc-pVTZ data that could successfully capture the dissociation process of CH$_5^+$ was reported~\cite{jin_ab_2006}, becoming the state-of-the-art PES of CH$_5^+$ since then.
Subsequently, researchers have applied various methods to investigate the (ro-)vibrational excited states and the infrared spectrum of CH$_5^+$ and its isotopologs.
These include large-scale vibrational configuration interaction (VCI) calculations~\cite{huang_deuteration_2006,huang_quantum_2006}, DMC methods~\cite{johnson_evolution_2006,hinkle_characterizing_2008,hinkle_theoretical_2009}, contracted basis-iterative method~\cite{wang_vibrational_2008,wang_computing_2015,wang_calculated_2016,fabri_use_2017}, (multilayer) multiconfigurational time-dependent Hartree method~\cite{wodraszka_ch5_2015}, the molecule superrotor model method~\cite{schmiedt_collective_2016,schmiedt_rotation-vibration_2017}, and the quantum-graph (QG) model~\cite{fabri_vibrational_2018,rawlinson_quantum_2019,rawlinson_exactly_2021,simko_quantum-chemical_2023}. 
Despite significant advances in experimental and computational methods, the accurate assignment of spectral lines for $\text{CH}_5^+$ remains an unsolved problem. Consequently, considerable effort is still required to interpret its spectrum and elucidate the molecule's complex dynamics.

\begin{figure}[htbp]
    \centering
    \includegraphics[width=\linewidth]{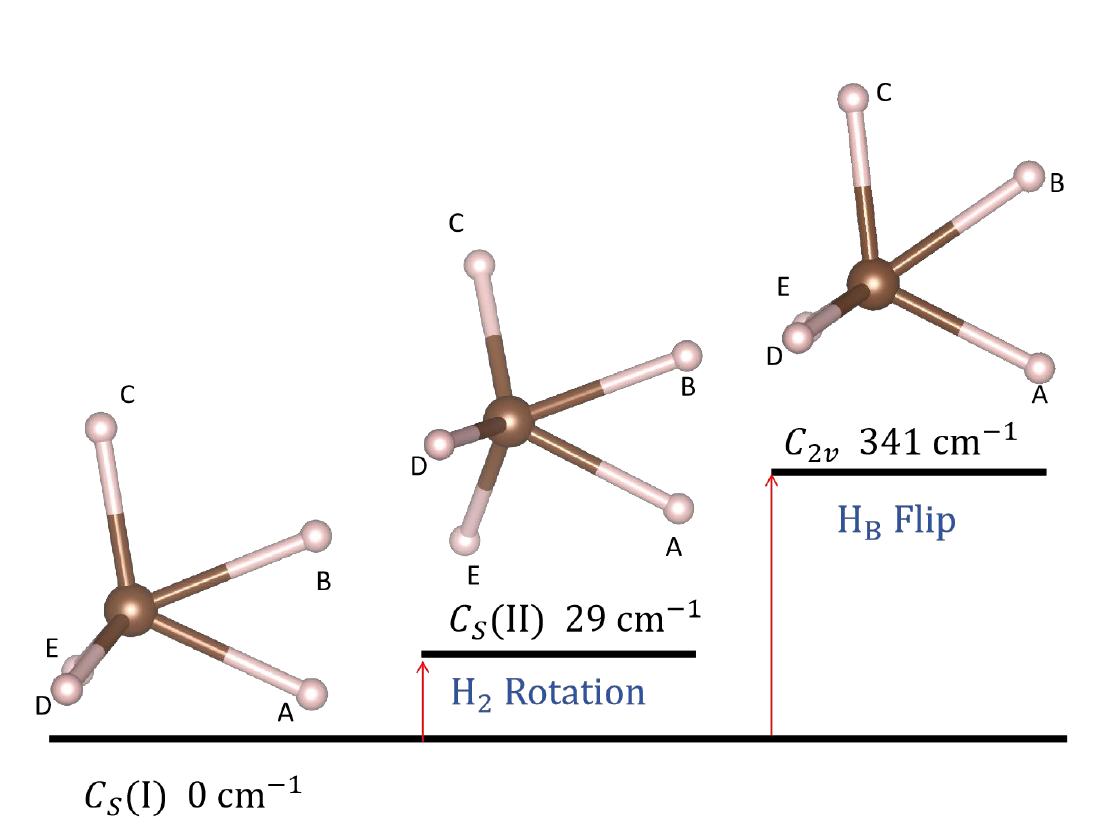}
    \caption{The structures of the three stationary points on the $\text{CH}_5^+$ PES with energies relative to the global minimum at the CCSD(T) level from Ref.~\cite{huang_quantum_2006}.
    In the $C_s\text{(I)}$ configuration, the hydrogens A, B, and C lie in the same plane as the central carbon.
    D and E are positioned symmetrically with respect to this plane. The $\text{H}_2$ moiety is formed by A and B. In the $C_s\text{(II)}$ configuration, the $\text{H}_2$ moiety (also formed by A and B) is rotated by 30$^\circ$ relative to the $C_s\text{(I)}$ configuration, and another 30$^\circ$ rotation of the H$_2$ moiety would lead back to another $C_s\text{(I)}$ configuration. In the $C_{2v}$ configuration, the hydrogens labeled A and C are symmetric with respect to the plane formed by D, B, E, and the central carbon.
    }
    \label{fig:three_configs}
\end{figure}

The addition of a proton to methane (CH$_4$) to form CH$_5^+$ causes a significant structural deviation from the tetrahedral geometry of methane.
The relatively rigid structure of $\text{CH}_4$ creates high energy barriers that prevent the free permutation of its hydrogen atoms during molecular vibration. This allows each hydrogen atom to be uniquely labeled.
An equilibrium geometry with an arbitrary but fixed numbering of the atoms can be chosen as a reference configuration. During vibration, the molecule primarily explores regions of the PES near this reference geometry~\cite{gray_anharmonic_1979,lee_accurate_1995}. Because the dynamics are localized in this manner, analyses based on normal coordinates are both appropriate and successful~\cite{wilson_molecular_1955,carter_variational_1999,carter_variational_2000}.
In contrast, the PES of CH$_5^+$ features 120 equivalent minima that differ from one another only by the permutation of hydrogen atoms. These minima are interconnected by low-energy barriers, causing the ground-state wavefunction to be delocalized over all 120 configurations.
Between any two equivalent global minima on the PES, there exist saddle points (transition states) of two distinct symmetries (see~\Fig{fig:three_configs}, visualized by VESTA~\cite{momma_vesta_2011}).
Starting from a global minimum with $C_s$(I) symmetry, a $C_s$(II) saddle point is accessed via a $\text{30}^\circ$ rotation of the $\text{H}_2$ moiety. Alternatively, a flip motion that exchanges hydrogens between the $\text{H}_2$ and the $\text{CH}_3$ components proceeds through a $C_{2v}$ saddle point.
Both saddle points represent low energy barriers relative to the global minimum: the barrier height is only 29~$\text{cm}^{-1}$ for the $C_s$(II) point and 341~$\text{cm}^{-1}$ for the $C_{2v}$ point.
The resulting fluxionality of $\text{CH}_5^+$ produces a complex energy level structure across its spectrum. A prime example is the 175~$\text{cm}^{-1}$ level, which is assigned to a highly anharmonic $\text{H}_\text{B}$ flip isomerization~\cite{huang_quantum_2006}—a large-amplitude motion with no counterpart in the spectrum of rigid CH$_4$.
Furthermore, approximately 900 spectral lines are observed in the 2770-3150~$\text{cm}^{-1}$ region, as reported in Ref.~\cite{white_ch_1999}, which significantly exceeds the number of spectral lines found in that region for $\text{CH}_4$ and $\text{CH}_3^+$~\cite{bowman_variational_2008}.

The applications of machine learning methods to the computation of the PES, molecular spectra, and accelerating quantum chemistry calculations have recently gained significant attention, with several fast-developing fields emerging in parallel. A substantial body of work learns the PES itself with a neural network trained on ab initio data, and then uses the surrogate PES inside a conventional solver~\cite{manzhos_neural_2021,manzhos_machine_2023,ren_machine_2021,han_concise_2022,shanavas_rasheeda_high-dimensional_2022,han_ai-powered_2025,ishii_development_2022}. For $\text{CH}_5^+$ in particular, a neural network PES fitted from a small DMC-generated dataset has been used to accelerate DMC ground-state calculations on GPUs~\cite{dirisio_gpu-accelerated_2021}, and molecular-orbital-based machine learning at CCSD(T) accuracy has been combined with a neural network PES to compute vibrational excited states~\cite{lu_fast_2022}. Generative normalizing flows have also been applied to molecular problems, for example to sample molecular equilibrium configurations~\cite{noe_boltzmann_2019} or to learn optimal vibrational coordinates of small molecules under the variational principle~\cite{saleh_computing_2025,10.1063/5.0285954}. Convergence properties of such normalizing-flow / Hermite-basis spectral expansions have recently been studied from an approximation-theoretic standpoint in Ref.~\cite{saleh_convergence_2026}. In parallel, operator-learning architectures such as the Fourier Neural Operator~\cite{li_fourier_2021,kovachki2023neural} have been proposed as efficient surrogates for a range of quantum-mechanical computations~\cite{mizera_scattering_2023,shah_fourier_2026,jin_v2rho-fno_2026}.
Among these methods, the neural canonical transformation (NCT), originally developed for classical Hamiltonian systems~\cite{li_neural_2020} and later extended to quantum many-body problems~\cite{hao_xie_ab-initio_2022}, is a notable variational approach for the scalable and simultaneous computation of multiple excited states. In the quantum setting, NCT parameterizes a unitary transformation on the Hilbert space of a given Hamiltonian: starting from an orthonormal basis of known reference wavefunctions, a normalizing flow~\cite{papamakarios_normalizing_2021,kobyzev_normalizing_2021} acting on the coordinates --- together with a Jacobian-determinant factor --- maps this basis to a new orthonormal basis that is variationally optimized against the target Hamiltonian. A key advantage of NCT is that it provides direct access to the wavefunctions, which simplifies the calculation of physical properties.
The NCT approach has been successfully applied to various systems, including the thermal properties of interacting fermions~\cite{hao_xie_ab-initio_2022}, the effective mass of interacting electrons~\cite{xie_m_2023}, the equation of state of dense hydrogen in extreme situations~\cite{li2025deepvariationalfreeenergy}, the quantum anharmonic effects of quantum solids~\cite{zhang_neural_2024-1}, the vibrational eigenstates of molecules~\cite{zhang_neural_2024} and the high-pressure water ice~\cite{zhang2025quantumanharmoniceffectshydrogenbond}.

The primary objective of this paper is to apply the NCT method to compute the ground and excited eigenstates of the highly fluxional molecule $\text{CH}_5^+$.
Our previous implementation of NCT for molecular vibrations~\cite{zhang_neural_2024} used normal coordinates, which are unsuitable for a highly fluxional system like $\text{CH}_5^+$, whose wavefunction is delocalized over many equivalent configurations~\cite{wang_vibrational_2008}. The NCT method itself is independent of the coordinate choice: it can be applied in any coordinate system as long as a suitable reference basis is available. To accurately model $\text{CH}_5^+$, we therefore work directly with the three-dimensional Cartesian coordinates of the atoms in the Eckart frame~\cite{eckart_studies_1935}, and apply NCT there to compute the eigenstates.
The code is available at \url{https://github.com/Callo42/nct-protonated-methane}.

\section{Methods}
\label{sec:methods}

This section is organized as follows. First, we set up the nuclear Schrödinger equation. Second, we detail the NCT technique, covering the basis set, the construction of the target wavefunctions, the energy estimation procedure, and the optimization of the wavefunction ansatz. Finally, we compare the NCT method with other computational approaches for $\text{CH}_5^+$ calculations.

\subsection{Hamiltonian}
\label{subsec:vib_hamiltonian}

In this work, we perform a $J{=}0$ pure vibrational calculation for the ground and excited states of CH$_5^+$ by solving the time-independent Schrödinger equation for the nuclear Hamiltonian. All atomic coordinates throughout this work are defined in the Eckart frame~\cite{eckart_studies_1935}, so that translational and rotational degrees of freedom have been removed. For conciseness, we denote the Eckart-fixed coordinates simply as $\vec{x}$.
Under the Born-Oppenheimer approximation, the potential energy of the molecule is given by a function of its nuclear coordinates $\vec{x}$.
Under these considerations, the Hamiltonian reads:
\begin{equation}
    H = - \sum_{i,\alpha} \frac{1}{2m_i} 
    \frac{\partial^2}{\partial x^2_{i\alpha}} + V(\vec{x}), 
    \label{eq:hamiltonian}
\end{equation} 
where $m_i$ is the mass of the $i$-th atom and $\alpha$ is summed over three spatial dimensions.
We use the ab initio-based, full-dimensional PES $V(\vec{x})$ that is fitted to 36,173 CCSD(T)/aug-cc-pVTZ calculations from Ref.~\cite{jin_ab_2006}.
As pointed out in Ref.~\cite{wang_computing_2015}, starting from normal coordinates is not sufficient for an accurate calculation of $\text{CH}_5^+$.
We confirm this with our own NCT ablation in normal coordinates, in which the variational energy eventually falls below the reference zero-point energy -- a manifest violation of the variational principle that originates from the fact that normal coordinates are defined with respect to a single reference configuration and describe only vibrations in its vicinity, so that the Monte Carlo sampler cannot maintain ergodicity across the full CH$_5^+$ hydrogen-permutation manifold.
Therefore, we switch the coordinate system from the normal coordinates in our previous work~\cite{zhang_neural_2024} to the three-dimensional Cartesian coordinates.

\subsection{Neural Canonical Transformation}
\label{subsec:nct}

\begin{figure}[htbp]
    \centering
    \includegraphics[width=\linewidth]{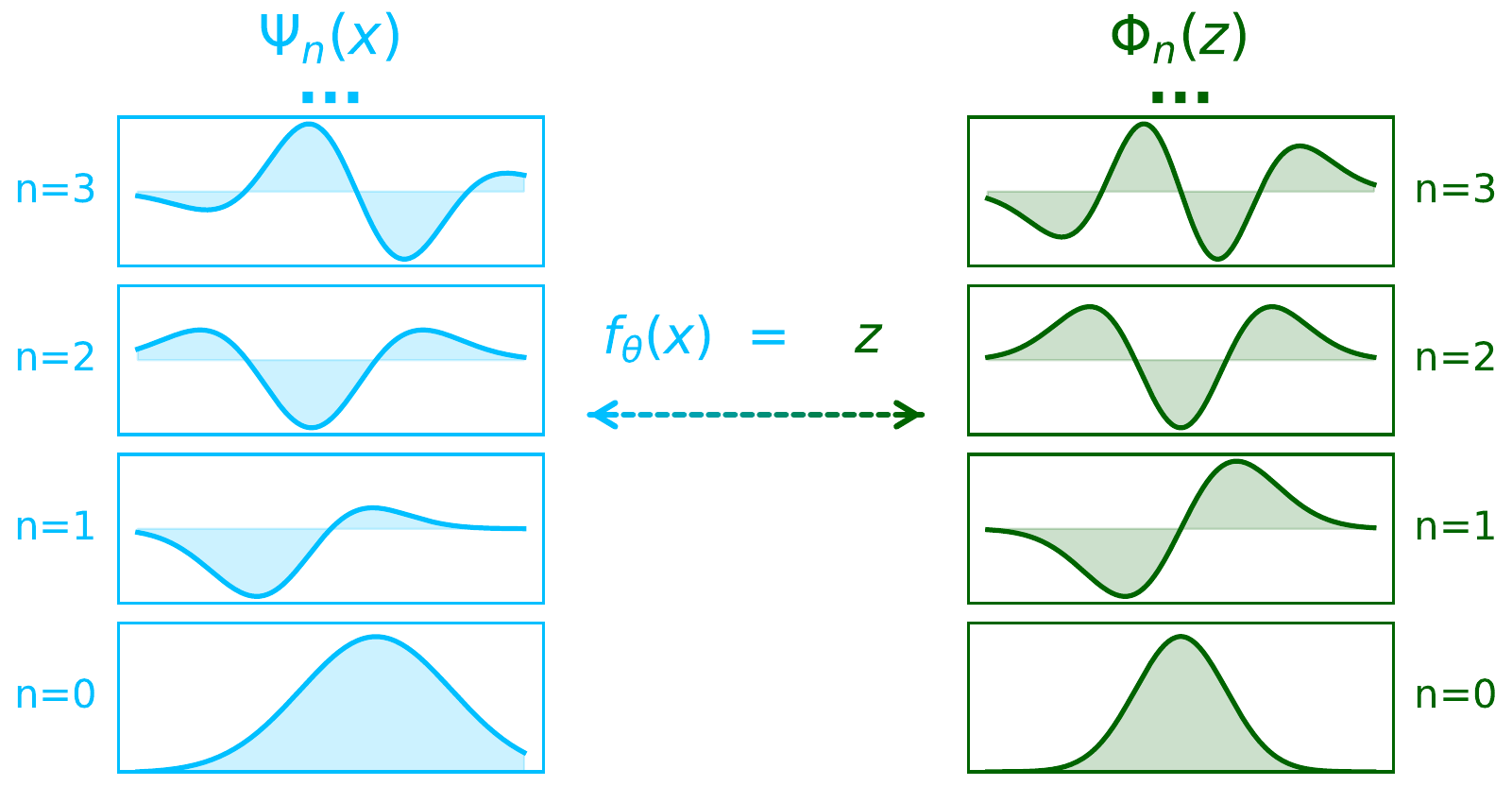}
    \caption{Neural Canonical Transformation: a sketch on a one-dimensional example. The $\Psi_n(x)$ in the left panel denotes the true wavefunction and the $\Phi_n(f_\theta(x))$ on the right refers to the wavefunction basis. $f_\theta(x)$ in the middle is the bijection parameterized by neural network that connects two sets of wavefunctions.
    }
    \label{fig:nct_sketch}
\end{figure}

In this study, we employ a normalizing flow~\cite{papamakarios_normalizing_2021,kobyzev_normalizing_2021} model to realize the NCT. This method performs a unitary transformation on an initial basis to yield wavefunctions that incorporate physical interactions, such as anharmonic effects~\cite{xie_m_2023}.
Consider the one-dimensional toy problem illustrated in~\Fig{fig:nct_sketch}. We begin with an orthonormal basis set composed of one-dimensional Hermite functions, $\left\{\Phi_n\right\}$, where $n$ is the excitation number and each function is an eigenstate of a one-dimensional harmonic oscillator (HO). A neural network then parameterizes a bijective mapping from this simple basis to the set of true, anharmonic wavefunctions, $\left\{\Psi_n\right\}$. This transformation is constructed to preserve orthonormality while simultaneously introducing anharmonic effects into the final wavefunctions.

The NCT method employed in this work uses a framework similar to our previous study~\cite{zhang_neural_2024}. However, to properly describe the fluxional nature of $\text{CH}_5^+$, we use Cartesian coordinates. The first step in calculating the eigenstates of the Hamiltonian in~\Eq{eq:hamiltonian} is to define the wavefunction basis, $\left\{\Phi_{\vec{n}}(\vec{z})\right\}$, in a latent space. This basis is constructed as the Hartree product of 18 one-dimensional HO wavefunctions:
\begin{equation}
    \Phi_{\vec{n}}(\vec{z}) = \prod_{i=1}^{18}  \phi_{n_i}(z_i),
    \label{eq:Phi_z}
\end{equation}
where 
\begin{equation}
        \phi_{n_i}(z_i)
    = \frac{1}{\sqrt{2^{n_i} n_i!}}  \left(\frac{m_i\omega_i}{\pi}\right)^{1/4}
         e^{\frac{-m_i\omega_i z_i^2}{2}}  H_{n_i}\left(\sqrt{m_i\omega_i}z_i\right),
    \label{eq:1d_ho_basis}
\end{equation}
here $\vec{n}=(n_1,n_2,\ldots,n_{18})$ is the vector of quantum numbers, and $\vec{z}=(z_1,z_2,\ldots,z_{18})$ represents the coordinates in the 18-dimensional latent space. 
Each dimension $i$ is associated with a mass $m_i$ and a harmonic frequency $\omega_i$. For our $\text{CH}_5^+$ calculation, the first three coordinates are assigned the mass of a carbon atom, while the remaining 15 are assigned the mass of a hydrogen atom. All calculations are performed in atomic units, with final results converted to other units, such as wavenumbers, where appropriate. In our setting, the variational calculation is carried out within an internal Hilbert subspace in which the overall translational and rotational motions are removed.

Each one-dimensional Hermite function is an eigenstate of a one-dimensional quantum harmonic oscillator, characterized by a center point that serves as a hyperparameter. Since the molecule has 18 vibrational degrees of freedom, the basis is built from 18 such one-dimensional functions. The center of each function is set to the corresponding coordinate value of the global minimum configuration of the PES, providing a good starting point for the variational optimization. The frequencies are set to treat identical atoms equally: the 15 oscillators corresponding to the hydrogen atoms are assigned a single frequency, $\omega_H$, and the three oscillators for the carbon atom are assigned a different frequency, $\omega_C$, where $\omega_C < \omega_H$. Since this initialization is in the latent space, it does not constrain the physical wavefunctions, which are determined by the learned normalizing flow. This construction defines our 18-dimensional wavefunction basis in the latent space.
The quantum number vectors $\vec{n}$ of the basis functions are generated as follows: we enumerate all 18-dimensional non-negative integer vectors and sort them first by the total number of excitation quanta $\sum_{i=1}^{18} n_i$ in ascending order, and then in reverse-lexicographic order within each group of equal total quanta. The first entry $\vec{n}=\vec{0}$ corresponds to the ground state. We then truncate the sequence to the lowest $N$ states.

Then a NCT is established to transform the wavefunction basis $\left\{\Phi_n(\vec{z})\right\}$ into the configuration space $\vec{x}$:
\begin{equation}
    \Psi_{\vec{n}}(\vec{x}) = \Phi_{\vec{n}}(f_\theta(\vec{x}))\left|\det {\left(\frac{\partial f_\theta (\vec{x})}{\partial \vec{x}}\right)} \right|^{1/2},
    \label{eq:nct_mapping}
\end{equation}
where $f_\theta \colon \mathbb{R}^{18} \to \mathbb{R}^{18}$ is an invertible mapping parameterized by a normalizing flow neural network with learnable parameters $\theta$.
The transformed wavefunctions $\left\{\Psi_{\vec{n}}(\vec{x})\right\}$ span a variational subspace of the Hamiltonian~\cite{courant_methods_2009}.
The combination of the coordinate map $f_\theta$ and the Jacobian-determinant factor in \Eq{eq:nct_mapping} defines a unitary transformation on the Hilbert space of the nuclear Hamiltonian: it takes the orthonormal basis $\left\{\Phi_{\vec{n}}\right\}$ to another orthonormal basis $\left\{\Psi_{\vec{n}}\right\}$, as expressed in \Eq{eq:orthonormality} below. This is the origin of the ``canonical transformation'' in the name NCT. The framework was originally introduced for classical Hamiltonian systems by Li et al.~\cite{li_neural_2020}, where ``canonical'' refers to a symplectic map on classical phase space parameterized by a normalizing flow with an explicit symplectic constraint. The analogous construction in quantum mechanics~\cite{hao_xie_ab-initio_2022} replaces the symplectic condition by unitarity on the Hilbert space; here the constraint is enforced automatically by the $|\det (\partial f_\theta/\partial \vec{x})|^{1/2}$ factor rather than by a dedicated architectural constraint on the flow. In both settings, the flow is constrained by the physics of the target Hamiltonian --- classical phase-space structure or quantum Hilbert-space structure --- which distinguishes NCT from other constrained normalizing flows.
In practice, before the coordinates are passed to the flow $f_\theta$, they are re-projected onto the Eckart frame to ensure translational and rotational invariance of the wavefunction.
During the transformation, it is expected that interactions such as collective dynamics of the atoms or anharmonicity are introduced into the wavefunction.
Here, we parameterize $f_\theta$ using real-valued non-volume-preserving (RNVP) transformations~\cite{dinh_density_2017,zhang_neural_2024}.
For conciseness, we abbreviate $\theta$ in the transformed wavefunction.
The square root of the Jacobian determinant term, $\left|\det {\left(\frac{\partial f_\theta (\vec{x})}{\partial \vec{x}}\right)} \right|^{1/2}$, is introduced by the variable substitution process in normalizing flow and is the key factor that ensures the orthonormality of the transformed wavefunctions~\cite{zhang_neural_2024}:
\begin{equation}
    \left\langle \Psi_{\vec{n}}|\Psi_{\vec{n'}}\right\rangle=\delta_{\vec{n}\vec{n'}}.
    \label{eq:orthonormality}
\end{equation}

Then we estimate the energy levels of the Hamiltonian in \Eq{eq:hamiltonian} as:
\begin{equation}
    E_{\vec{n}} =     \left\langle \Psi_{\vec{n}}|H|\Psi_{\vec{n}}\right\rangle = \underset{\vec{x} \sim |\Psi_{\vec{n}}(\vec{x})|^2}{\mathbb{E}} \left[ E_{\vec{n}}^{\text{loc}}(\vec{x}) \right],
    \label{eq:energy_eigenvalue}
\end{equation} 
here, $\mathbb{E}[\cdot]$ denotes the expectation of the quantity inside the square brackets. The term $\left\langle \Psi_{\vec{n}}|H|\Psi_{\vec{n}}\right\rangle$ is the expectation value of the Hamiltonian for the eigenstate $\Psi_{\vec{n}}$. Finally, $E_{\vec{n}}^{\text{loc}}(\vec{x})$ denotes the local energy of the wavefunction~\cite{zhang_neural_2024}:
\begin{equation}
    E_{\vec{n}}^{\text{loc}}(\vec{x}) = \sum_{i,\alpha} -\frac{1}{2m_i}\left[ \frac{\partial^2}{\partial x_{i\alpha}^2} \ln |\Psi_{\vec{n}}(\vec{x})| + \left( \frac{\partial}{\partial x_{i\alpha}} \ln |\Psi_{\vec{n}}(\vec{x})| \right)^2 \right] + V(\vec{x}).
    \label{eq:local_energy}
\end{equation}
With access to the wavefunctions in \Eq{eq:nct_mapping}, the energies in \Eq{eq:energy_eigenvalue} are estimated by Markov chain Monte Carlo (MCMC)~\cite{becca_quantum_2017} sampling of the wavefunction.
To restrict sampling to the $J{=}0$ pure vibrational subspace, the MCMC proposals are projected onto the Eckart frame, removing translational and rotational components from each proposed displacement (see the Supplementary Information for details).

To optimize the neural network parameters, we employ the ensemble Rayleigh--Ritz variational principle~\cite{gross_rayleigh-ritz_1988,courant_methods_2009,zhang_neural_2024} and define the loss function as a Boltzmann-weighted sum of state energies:
\begin{equation}
    \mathcal{L} = \sum_{\left\{\vec{n}\right\}} w_{\vec{n}} \, E_{\vec{n}},
    \label{eq:loss}
\end{equation}
where the weights $w_{\vec{n}}$ are fixed Boltzmann factors at a chosen temperature $T$, determined by the harmonic-approximation vibrational energy levels $\varepsilon^{\mathrm{harm}}_{\vec{n}}$ obtained directly from the PES:
\begin{equation}
    w_{\vec{n}} = \frac{e^{-\beta \varepsilon^{\mathrm{harm}}_{\vec{n}}}}{\sum_{\left\{\vec{m}\right\}} e^{-\beta \varepsilon^{\mathrm{harm}}_{\vec{m}}}}, \qquad \beta = \frac{1}{k_B T}.
    \label{eq:boltzmann_weights}
\end{equation}
The harmonic levels $\varepsilon^{\mathrm{harm}}_{\vec{n}}$ are chosen as the lowest-lying levels up to the total number of states computed; they serve solely to set the weights and do not have a one-to-one correspondence with the actual excitation indices of the optimized wavefunctions.
The Boltzmann weighting naturally assigns larger importance to lower-energy states, and minimizing $\mathcal{L}$ yields the optimal variational estimate of the wavefunctions within the transformed subspace. Under this principle, once the optimal variational estimate is reached, each individual energy level $E_{\vec{n}}$ attains its best approximation simultaneously.
In practice, we simultaneously solve for dozens of energy levels, and consequently, the summation in \Eq{eq:loss} typically involves a similar number of terms.

The gradient of the loss function with respect to the neural network parameters is estimated as~\cite{zhang_neural_2024,hao_xie_ab-initio_2022,xie_m_2023}:
\begin{equation}
    \nabla_{\theta} \mathcal{L} = 2 \sum_{\left\{\vec{n}\right\}} w_{\vec{n}} \underset{\vec{x} \sim |\Psi_{\vec{n}}(\vec{x})|^2}{\mathbb{E}} \left[ E_{\vec{n}}^{\text{loc}}(\vec{x}) \nabla_{\theta} \ln |\Psi_{\vec{n}}(\vec{x})| \right].
    \label{eq:loss_gradient}
\end{equation}
In \Eq{eq:loss_gradient}, the expectation is estimated by the same MCMC sampling used for the energy in \Eq{eq:energy_eigenvalue}.
This gradient estimator takes the form of the REINFORCE gradient~\cite{williams_simple_1992} and has been shown to be stable in variational Monte Carlo settings with normalizing flows~\cite{zhang_neural_2024,hao_xie_ab-initio_2022}.
We use the Adam~\cite{kingma_adam_2017} optimizer to perform gradient descent on the parameters.
After convergence is reached, each energy level is measured following \Eq{eq:energy_eigenvalue}.

\subsection{Comparison to Other Methods}
\label{subsec:compare_other_method}

It is also useful to position NCT against the machine learning approaches surveyed in Sec.~\ref{sec:intro}. The neural network PES methods learn the potential energy surface itself from ab initio data and then feed it into a conventional solver for the Schr\"odinger equation; in contrast, NCT learns the wavefunctions directly and is in principle agnostic to how the PES is provided, so that an NN-PES surrogate could in fact be plugged into the NCT workflow without changing the variational framework. The flow-based molecular models discussed in Sec.~\ref{sec:intro} use a normalizing flow either to sample molecular equilibrium configurations or to learn optimal vibrational coordinates inside a fixed wavefunction basis, whereas in NCT the normalizing flow plays a qualitatively different role: together with the Jacobian-determinant factor, it parameterizes a unitary transformation on the Hilbert space of the nuclear Hamiltonian, mapping a tractable orthonormal basis into a variationally optimized eigenbasis.

The theoretical methods for $\text{CH}_5^+$ that have been discussed in Sec.~\ref{sec:intro} are valuable approaches that have had success in different areas. However, they still encounter various challenges.
For instance, methods like VCI and iterative eigensolvers tend to need large basis sets for convergence, which can make scaling up the calculations to more excited states quite challenging.
The superrotor and QG models, on the other hand, introduce several approximations to the PES, and combining different PES into the model can require considerable effort.
DMC methods require manually constructing nodal surfaces under the fixed-node approximation, a process that depends on prior knowledge of the specific energy level. This dependence poses challenges for simultaneously computing multiple excited states, which is particularly important in the case of $\text{CH}_5^+$. 
Compared to existing computational methods for the spectra of $\text{CH}_5^+$, the NCT method stands out as an ab initio approach that readily and systematically scales to a large number of excited states and is able to seamlessly incorporate various PES.

\section{Results}
\label{sec:results}

This section is organized as follows. We begin in Sec.~\ref{subsec:num_detail} by describing the numerical details of our calculations. In Sec.~\ref{zpe_rdf_gs}, we analyze the zero-point energy and radial distribution functions of the ground state. Next, Sec.~\ref{subsec:excite_lorentz} presents the Lorentzian spectrum derived from the computed energy levels. Finally, Sec.~\ref{subsec:relative_similarity} reports our findings on relative similarity for various states.

\subsection{Numerical Details}
\label{subsec:num_detail}

In this work, we carried out two calculations: the first involves a single calculation of the ground state, while the second targets the lowest 32 vibrational eigenstates (ground state plus 31 excited states) of the CH$_5^+$ nuclear Hamiltonian. 
All calculations were performed in atomic units for a system of one carbon and five hydrogen atoms (see Sec.~\ref{subsec:nct}).
For the ground-state calculation, the RNVP~\cite{zhang_neural_2024,dinh_density_2017} network is initialized with 316{,}082 learnable parameters, and the Adam~\cite{kingma_adam_2017} optimizer is set with a learning rate of $3 \times 10^{-4}$; we use a total batch size of 6000 for steady convergence behavior and optimize for 20{,}000 iterations.
For the excited-state calculation, the RNVP network is initialized with 88{,}754 learnable parameters, and the Adam optimizer is set with a learning rate of $5 \times 10^{-5}$; we train with 1200 samples per energy level and optimize for 10{,}000 iterations.
The narrower network for the excited-state run is a deliberate trade-off: since all 32 states share a single flow whose energy and gradient must be evaluated at every iteration, a narrower network lets us afford a sufficient per-state batch at a total sampling cost comparable to the ground-state run; precision is generally expected to improve with network capacity, and the current network width is the best trade-off we have found for this study. As shown in Table~\ref{tab:full_energy}, the resulting excited-state energies are already comparable to the reference variational and DMC results on the same PES.
For energy estimation of the ground state, we make an extra 100 iterations of Monte Carlo (MC) sampling of the converged wavefunctions, resulting in the equivalent of 600,000 samples. The zero-point energy is directly estimated on the sampled wavefunction.
For excited states, we perform an extra 100 iterations of Monte Carlo sampling (equivalent to 120,000 samples per state) on the converged wavefunctions, and all eigenstates included in training are sampled simultaneously. 
For the radial distribution functions of C-H and H-H distances and the relative-similarity analysis to the three stationary points, we use the final converged checkpoint and sample each wavefunction individually.
The ground-state calculation took approximately $3$~hours on a compute node with $4\times$ NVIDIA RTX~5090 GPUs, and the $32$-state excited-state calculation took approximately $6$~hours $14$~minutes on a compute node with $4\times$ NVIDIA A100~$80$\,GB GPUs, with the batch distributed across devices via \texttt{jax.pmap}. A detailed measurement of the per-iteration wall-clock scaling with the number of simultaneously optimized states is reported in Appendix~\ref{subsec:comp_cost}.

\subsection{Zero-point Energy and Radial Distribution Functions of Ground State}
\label{zpe_rdf_gs}

The zero-point energy we obtain is 10917.98(2.90)~$\text{cm}^{-1}$, which agrees with the reference zero-point energy reported in the PES paper~\cite{jin_ab_2006}. This agreement validates the adequacy of our computational setup for capturing the unique properties of this molecule. Without further architectural modification, we proceed to scale our calculation to excited states.
In the following, we will illustrate that the NCT method successfully learns a delocalized wavefunction that explores the equivalent minima on the PES, capturing the fluxional nature of CH$_5^+$.

\begin{figure}[htbp]
    \centering
    \includegraphics[width=\linewidth]{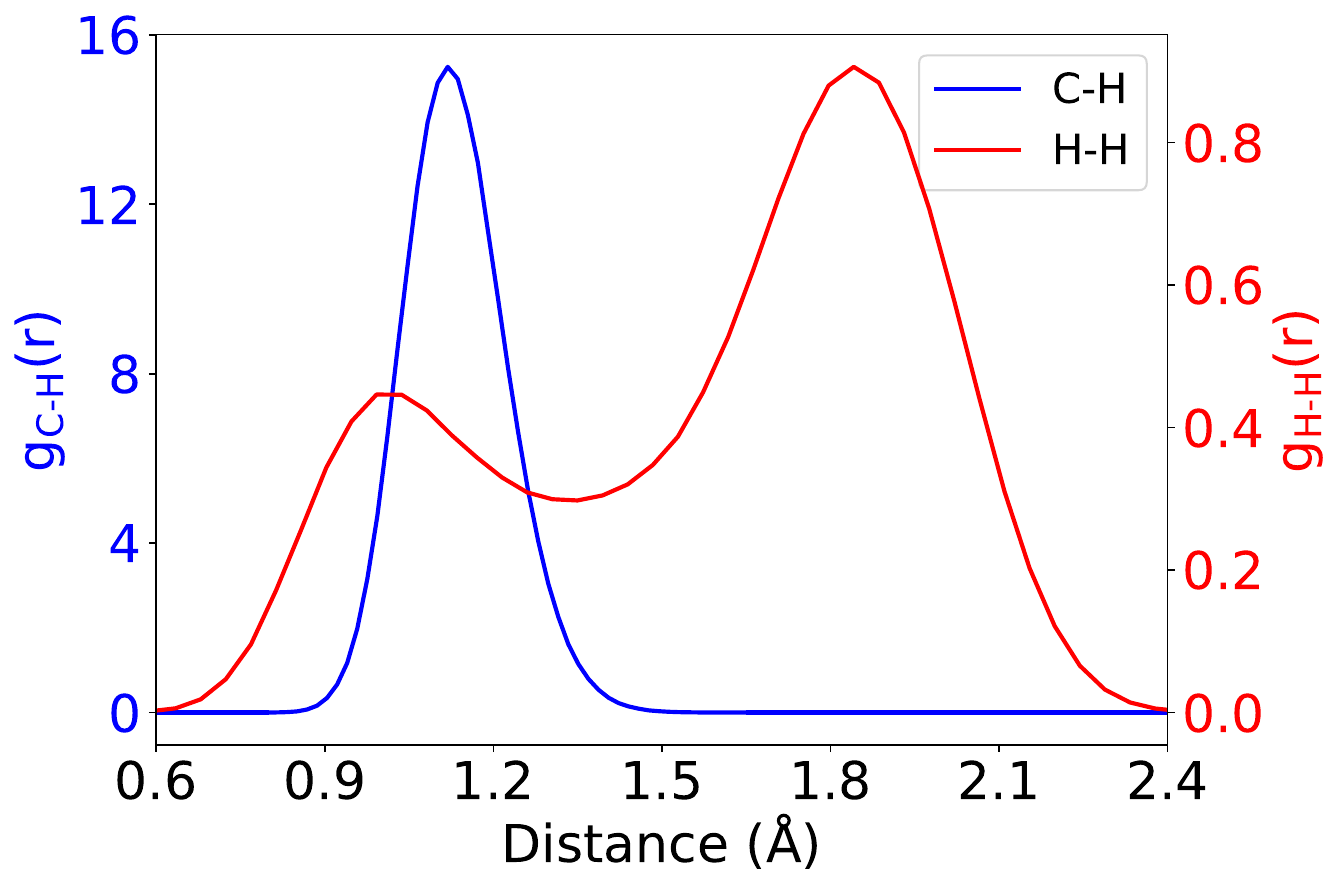}
    \caption{The radial distribution functions 
    for C-H distances (in angstroms, left axis) and H-H distances (in angstrom, right axis) of the converged NCT wavefunction of ground state CH$_5^+$.
    }
    \label{fig:gs_ch_hh}
\end{figure}

The radial distribution functions of C-H distances and H-H distances of the ground state are shown in \Fig{fig:gs_ch_hh}. 
The units of the C-H distances and the H-H distances are given in angstroms.
The structure of CH$_5^+$ is learned by the wavefunction as the NCT method explores the PES.

The results show that the C-H distances form a wide unimodal distribution. This indicates that all the C-H bonds have significant flexibility in length, making it impossible to distinguish individual C-H bonds by bond length alone.
The H-H distances form a bimodal distribution. The first peak occurs around 1.0~\AA~with weak intensity, and the main peak takes place around 1.9~\AA. The 1.0~\AA~peak indicates that there are a significant number of hydrogen pairs that are close to each other, forming the H$_2$ moiety. The 1.9~\AA~peak is formed by other hydrogen pairs. The unimodal C-H distances and bimodal H-H distances confirm that the quantum ground state is dominated by CH$_3^+$ and H$_2$, which is also consistent with previous studies in Ref.~\cite{thompson_ch_2005,marx_structural_1995,hinkle_characterizing_2008}.

\subsection{Excited States and Lorentzian Spectrum}
\label{subsec:excite_lorentz}

The excited-state energy levels from the 32-state CH$_5^+$ calculation are listed in Table.~\ref{tab:full_energy}, in which the zero-point energy is omitted and the 31 excited states are reported as relative energy differences to the zero-point energy in $\text{cm}^{-1}$.
In addition, we also calculate the lowest 24 excited energy levels of $\text{CH}_4$ using the NCT method with the PES from Ref.~\cite{lee_accurate_1995}. We have verified that performing the $\text{CH}_4$ calculation in the Eckart frame with Cartesian coordinates yields results consistent with those obtained using normal coordinates. Since the normal-coordinate formulation is the standard approach for $\text{CH}_4$, we present only the normal-coordinate results here, following the same scheme in Ref.~\cite{zhang_neural_2024}.

\begin{figure}[htbp]
    \centering
    \includegraphics[width=\linewidth]{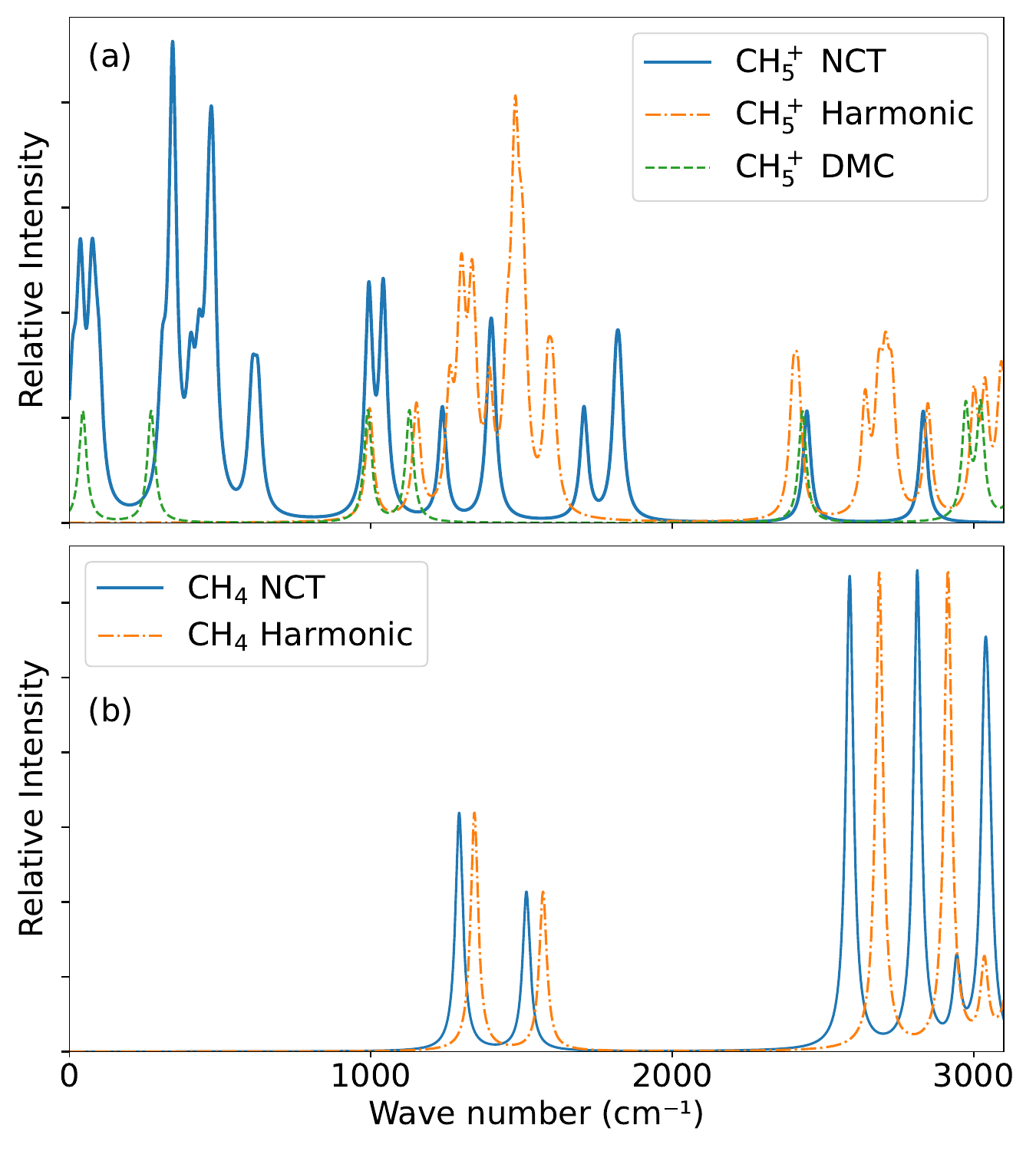}
    \caption{The Lorentzian spectra of the energy levels (convolved with FWHM of 30~$\text{cm}^{-1}$) for (a) $\text{CH}_5^+$ and (b) $\text{CH}_4$ from the NCT calculation and their harmonic approximations. For $\text{CH}_5^+$, panel (a) also shows the fixed-node DMC excited-state energies of Ref.~\cite{hinkle_characterizing_2008} convolved with the same Lorentzian.
    The harmonic approximation result of $\text{CH}_5^+$ is from the normal mode frequencies in Ref.~\cite{jin_ab_2006}. The harmonic results for $\text{CH}_4$ are from the PES in Ref.~\cite{lee_accurate_1995}.
    }
    \label{fig:lorentz_spectra}
\end{figure}

In contrast to the IR spectrum reported in Ref.~\cite{huang_quantum_2006}, which includes broadening from IR intensities, our approach generates a spectrum by directly broadening the calculated energy levels with equal weight for each level. To achieve this, we apply a Lorentzian convolution with a full width at half maximum of 30~$\text{cm}^{-1}$ to both our calculated NCT energy levels and the levels from harmonic approximations, as shown in \Fig{fig:lorentz_spectra}.
The harmonic approximation results for $\text{CH}_5^+$ are from the normal mode frequencies in Ref.~\cite{jin_ab_2006} and include only fundamental levels above 900~$\text{cm}^{-1}$ because the low-frequency levels are highly anharmonic~\cite{huang_quantum_2006}. The harmonic results for $\text{CH}_4$ are from the harmonic approximation of the PES in Ref.~\cite{lee_accurate_1995}.

Before discussing the specific differences in Fig.~\ref{fig:lorentz_spectra}, we emphasize that the NCT and harmonic curves have fundamentally different status. The NCT spectrum is obtained by variational optimization against the full anharmonic nuclear Hamiltonian of \Eq{eq:hamiltonian} under the Rayleigh--Ritz principle, and each NCT eigenvalue is therefore a variational estimate of an exact eigenvalue of the full Hamiltonian. The harmonic spectrum, by contrast, is a local second-order expansion and does not correspond to any eigenvalue of the true Hamiltonian. For CH$_5^+$, the harmonic approximation is known to break down severely because the molecule is highly fluxional and its wavefunction is delocalized over 120 equivalent minima connected by very low saddle points ($C_s$(II) at 29~cm$^{-1}$, $C_{2v}$ at 341~cm$^{-1}$); the NCT curve in Fig.~\ref{fig:lorentz_spectra}(a) is thus the more physically correct description, and the harmonic curve is shown only as a familiar reference point.

In comparison to the harmonic approximation of CH$_5^+$, the NCT Lorentzian spectrum displays notable discrepancies in several frequency regions. At low frequencies, we find several excited states with the lowest at about 10~$\text{cm}^{-1}$ that are completely absent from the harmonic approximation. These low-lying levels are in quantitative agreement with the converged full-dimensional variational calculation of Wang \& Carrington~\cite{wang_vibrational_2008} on the same JBB PES (see the state-by-state comparison in Table~\ref{tab:full_energy}), and the fixed-node DMC of Hinkle \& McCoy~\cite{hinkle_characterizing_2008} independently reports a state at 44~$\text{cm}^{-1}$ in the same region (see Fig.~\ref{fig:lorentz_spectra}(a)), confirming that they are genuine anharmonic features of the vibrational Hamiltonian rather than a method artifact and providing further evidence that the harmonic approximation fails for CH$_5^+$. These low-energy excitations are generally absent from experimental infrared spectra due to limited sensitivity in this frequency range~\cite{asvany_understanding_2005}, but they are present in the spectrum of the true vibrational Hamiltonian.
Near 2000~$\text{cm}^{-1}$, the NCT spectrum has additional energy levels that are also absent from the harmonic approximation. The DMC method can identify individual states in this region when the nodal surface is chosen properly; for example, Ref.~\cite{hinkle_theoretical_2009} reports a high-symmetry state at 2164~$\text{cm}^{-1}$ from DMC, although such a high-symmetry state is generally forbidden or carries negligible oscillator strength in an infrared spectrum.

The CH$_4$ panel of Fig.~\ref{fig:lorentz_spectra}(b) serves as a sanity check. In CH$_4$, atomic vibrations are confined near a single rigid equilibrium configuration protected by high barriers, giving sharp and isolated normal-mode peaks. The NCT result for CH$_4$ manifests primarily as small red shifts of the harmonic energy levels, in agreement with the findings of Ref.~\cite{lee_accurate_1995} --- i.e., when the harmonic approximation is nearly correct, NCT reproduces it up to residual anharmonicity. The contrast between this near-agreement for CH$_4$ and the dramatic NCT$\leftrightarrow$harmonic discrepancy for CH$_5^+$ is a direct visual signature of the unique complexity and fluxionality of CH$_5^+$: the same method gives small corrections for a rigid molecule and large corrections for a fluxional one, exactly as one would expect.
We will illustrate later that the wavefunctions of both the low-energy and high-energy excited states exhibit significant delocalization over the equivalent minima on the PES.

\subsection{Relative Similarity to three Stationary Points}
\label{subsec:relative_similarity}

\begin{figure}[htbp]
    \centering
    \includegraphics[width=\linewidth]{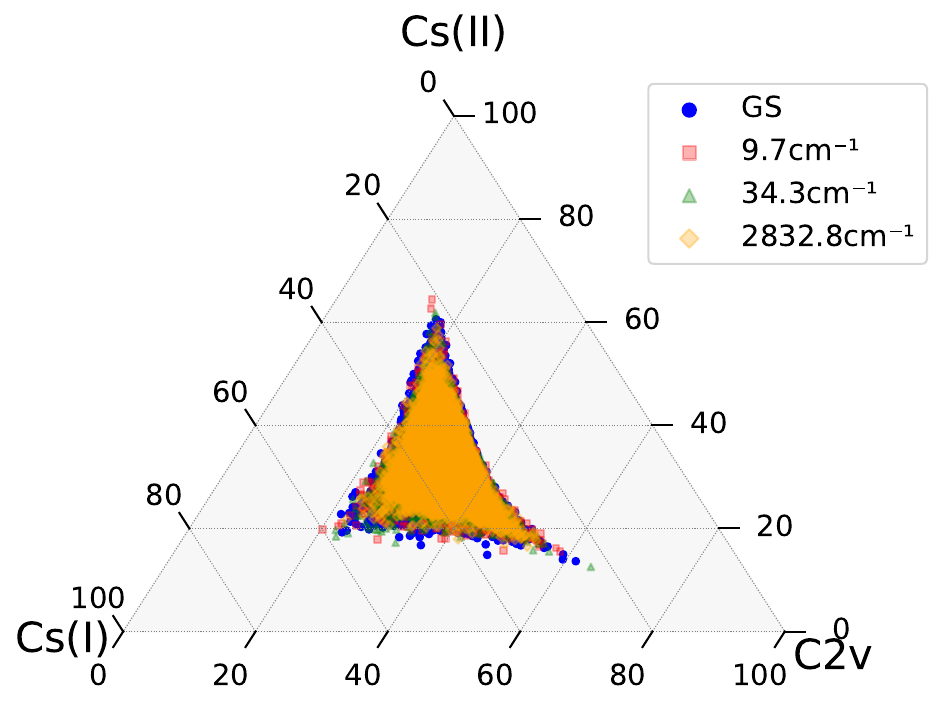}
    \caption{Ternary diagram of the relative similarity of samples from each state to the three types of stationary point configurations on PES. The values are percentages and are normalized for each sample. 
    }
    \label{fig:ternary_combined}
\end{figure}

For each chosen state, we first sample the converged wavefunction to generate 120,000 samples. For each sample $\vec x$ with atomic positions $\{\vec x_i\}_{i=1}^{N_{\mathrm{atom}}}$ and each reference configuration $k \in \{C_s(\mathrm{I}), C_s(\mathrm{II}), C_{2v}\}$, we compute the root mean square deviation $d_k(\vec x) = \sqrt{(1/N_{\mathrm{atom}})\sum_{i=1}^{N_{\mathrm{atom}}}\|\vec x_i - \vec x^{(k)}_i\|^2}$ between $\vec x$ and the reference geometry $\vec x^{(k)}$ (after optimal rigid-body alignment and hydrogen relabeling), define the similarity score $s_k(\vec x) = 1/d_k(\vec x)$, and take the normalized triple $t_k(\vec x) = s_k(\vec x) / \sum_{k'} s_{k'}(\vec x)$ as the barycentric coordinate of the sample in a ternary diagram whose three vertices are labeled by the three reference configurations, as illustrated in \Fig{fig:ternary_combined}.
For illustration, we choose the ground state, the first excited state at 9.7~$\text{cm}^{-1}$, and two additional excited states with energy levels of 34.3~$\text{cm}^{-1}$ and 2832.8~$\text{cm}^{-1}$.

The ground-state and excited-state distributions exhibit a common feature: both are centered and form three peaks directed toward the stationary points. This structure indicates a balanced preference for the $C_s$(I), $C_s$(II) and $C_{2v}$ stationary points. Because the molecule in these states has a similar preference for each configuration, there is a significant probability of finding it in any of them. This probability signifies a delocalized wavefunction, which in turn allows the molecule to move among these three configurations.

\section{Discussions}
\label{sec:discussion}

Strong anharmonicity and large-amplitude motions of hydrogen atoms in CH$_5^+$ give rise to numerous low-energy excitations, which constitute the most distinctive spectral feature of this molecule.
The NCT wavefunction is capable of expressing the delocalization of the nuclear wavefunction, thereby capturing the fluxional nature of CH$_5^+$.
We find that the nuclear ground state, as well as the low-lying and high-energy excited states, are highly anharmonic and exhibit similar preferences for the three stationary points on the PES.

This work extends the applications of the NCT method to fluxional molecules, enabling it to solve eigenstates with delocalized wavefunctions for systems that undergo large-amplitude spatial motion, such as CH$_5^+$.
Using the NCT method, we present a variational approach that solves the nuclear Schrödinger equation for CH$_5^+$. The method scales linearly with the number of excited states.
Finally, in future studies, we expect to enforce the correct hydrogen-permutation symmetry on the wavefunction to obtain more accurate energy levels.
A permutation-equivariant implementation of the flow model is necessary to achieve this, as demonstrated in Ref.~\cite{wirnsberger_targeted_2020}.
In particular, we expect several of the near-degenerate states presently observed in our NCT spectrum (Table~\ref{tab:full_energy}) to become strictly degenerate under such a symmetrized ansatz.

\begin{acknowledgments}
We gratefully thank Tucker Carrington Jr. for providing the potential energy surface of protonated methane. We are grateful for the useful discussions with Hao Xie, Xinguo Ren, Xin-Yang Dong, Lu Zhao, Zi-Hang Li, Zhen-Dong Cao, Shi-Gang Ou and Bei Qiao. This work is supported by the National Natural Science Foundation of China under Grants No. 92270107, No. T2225018, No. 12188101, No. T2121001 and the Strategic Priority Research Program of the Chinese Academy of Sciences under Grants No. XDB0500000, and the National Key Projects for Research and Development of China Grants No. 2021YFA1400400.

\end{acknowledgments}

\section*{Data Availability Statement}
The data and code that support the findings of this study are openly available at \url{https://github.com/Callo42/nct-protonated-methane}.

\bibliographystyle{apsrev4-2}
\bibliography{refs}

@article{carter_variational_2000,
	title = {Variational {Calculations} of {Rotational}−{Vibrational} {Energies} of {CH} $_{\textrm{4}}$ and {Isotopomers} {Using} an {Adjusted} ab {Initio} {Potential}},
	volume = {104},
	issn = {1089-5639, 1520-5215},
	url = {https://pubs.acs.org/doi/10.1021/jp991723b},
	doi = {10.1021/jp991723b},
	
	number = {11},
	urldate = {2024-05-09},
	journal = {The Journal of Physical Chemistry A},
	author = {Carter, Stuart and Bowman, Joel M.},
	month = mar,
	year = {2000},
	pages = {2355--2361},
}

@article{carter_variational_1999,
	title = {Variational calculations of rovibrational energies of {CH4} and isotopomers in full dimensionality using an \textit{ab initio} potential},
	volume = {110},
	issn = {0021-9606, 1089-7690},
	url = {https://pubs.aip.org/jcp/article/110/17/8417/475368/Variational-calculations-of-rovibrational-energies},
	doi = {10.1063/1.478750},
	
	number = {17},
	urldate = {2024-05-09},
	journal = {The Journal of Chemical Physics},
	author = {Carter, Stuart and Shnider, Heather M. and Bowman, Joel M.},
	month = may,
	year = {1999},
	pages = {8417--8423},
}

@article{dirisio_gpu-accelerated_2021,
	title = {{GPU}-{Accelerated} {Neural} {Network} {Potential} {Energy} {Surfaces} for {Diffusion} {Monte} {Carlo}},
	volume = {125},
	copyright = {https://doi.org/10.15223/policy-029},
	issn = {1089-5639, 1520-5215},
	url = {https://pubs.acs.org/doi/10.1021/acs.jpca.1c03709},
	doi = {10.1021/acs.jpca.1c03709},
	
	number = {26},
	urldate = {2024-04-07},
	journal = {The Journal of Physical Chemistry A},
	author = {DiRisio, Ryan J. and Lu, Fenris and McCoy, Anne B.},
	month = jul,
	year = {2021},
	pages = {5849--5859},
}

@article{lu_fast_2022,
	title = {Fast {Near} \textit{{Ab} {Initio}} {Potential} {Energy} {Surfaces} {Using} {Machine} {Learning}},
	volume = {126},
	copyright = {https://doi.org/10.15223/policy-029},
	issn = {1089-5639, 1520-5215},
	url = {https://pubs.acs.org/doi/10.1021/acs.jpca.2c02243},
	doi = {10.1021/acs.jpca.2c02243},
	
	number = {25},
	urldate = {2024-04-07},
	journal = {The Journal of Physical Chemistry A},
	author = {Lu, Fenris and Cheng, Lixue and DiRisio, Ryan J. and Finney, Jacob M. and Boyer, Mark A. and Moonkaen, Pattarapon and Sun, Jiace and Lee, Sebastian J. R. and Deustua, J. Emiliano and Miller, Thomas F. and McCoy, Anne B.},
	month = jun,
	year = {2022},
	pages = {4013--4024},
}

@article{gray_anharmonic_1979,
	title = {The anharmonic force field and equilibrium structure of methane},
	volume = {37},
	issn = {0026-8976, 1362-3028},
	url = {http://www.tandfonline.com/doi/abs/10.1080/00268977900101401},
	doi = {10.1080/00268977900101401},
	
	number = {6},
	urldate = {2024-04-02},
	journal = {Molecular Physics},
	author = {Gray, D.L. and Robiette, A.G.},
	month = jun,
	year = {1979},
	pages = {1901--1920},
}

@article{asvany_understanding_2005,
	title = {Understanding the {Infrared} {Spectrum} of {Bare} {CH} $_{\textrm{5}}$ $^{\textrm{+}}$},
	volume = {309},
	issn = {0036-8075, 1095-9203},
	url = {https://www.science.org/doi/10.1126/science.1113729},
	doi = {10.1126/science.1113729},
	
	number = {5738},
	urldate = {2024-03-27},
	journal = {Science},
	author = {Asvany, Oskar and P, Padma Kumar and Redlich, Britta and Hegemann, Ilka and Schlemmer, Stephan and Marx, Dominik},
	month = aug,
	year = {2005},
	pages = {1219--1222},
}

@article{oka_taming_2015,
	title = {Taming {CH} $_{\textrm{5}}$ $^{\textrm{+}}$ , the “enfant terrible” of chemical structures},
	volume = {347},
	issn = {0036-8075, 1095-9203},
	url = {https://www.science.org/doi/10.1126/science.aaa6935},
	doi = {10.1126/science.aaa6935},
	
	number = {6228},
	urldate = {2024-03-27},
	journal = {Science},
	author = {Oka, Takeshi},
	month = mar,
	year = {2015},
	pages = {1313--1314},
}

@article{huang_quantum_2006,
	title = {Quantum {Deconstruction} of the {Infrared} {Spectrum} of {CH} $_{\textrm{5}}$ $^{\textrm{+}}$},
	volume = {311},
	issn = {0036-8075, 1095-9203},
	url = {https://www.science.org/doi/10.1126/science.1121166},
	doi = {10.1126/science.1121166},
	
	number = {5757},
	urldate = {2024-03-27},
	journal = {Science},
	author = {Huang, Xinchuan and McCoy, Anne B. and Bowman, Joel M. and Johnson, Lindsay M. and Savage, Chandra and Dong, Feng and Nesbitt, David J.},
	month = jan,
	year = {2006},
	pages = {60--63},
}

@article{lee_accurate_1995,
	title = {An accurate \textit{ab} \textit{initio} quartic force field and vibrational frequencies for {CH4} and isotopomers},
	volume = {102},
	issn = {0021-9606, 1089-7690},
	url = {https://pubs.aip.org/jcp/article/102/1/254/482039/An-accurate-ab-initio-quartic-force-field-and},
	doi = {10.1063/1.469398},
	
	number = {1},
	urldate = {2024-03-25},
	journal = {The Journal of Chemical Physics},
	author = {Lee, Timothy J. and Martin, Jan M. L. and Taylor, Peter R.},
	month = jan,
	year = {1995},
	pages = {254--261},
}

@article{wilson_molecular_1955,
	title = {Molecular {Vibrations}: {The} {Theory} of {Infrared} and {Raman} {Vibrational} {Spectra}},
	volume = {102},
	issn = {00134651},
	shorttitle = {Molecular {Vibrations}},
	url = {https://iopscience.iop.org/article/10.1149/1.2430134},
	doi = {10.1149/1.2430134},
	
	number = {9},
	urldate = {2024-03-24},
	journal = {Journal of The Electrochemical Society},
	author = {Wilson, E. B. and Decius, J. C. and Cross, P. C. and Sundheim, Benson R.},
	year = {1955},
	pages = {235C},
}

@article{bowman_variational_2008,
	title = {Variational quantum approaches for computing vibrational energies of polyatomic molecules},
	volume = {106},
	issn = {0026-8976, 1362-3028},
	url = {http://www.tandfonline.com/doi/abs/10.1080/00268970802258609},
	doi = {10.1080/00268970802258609},
	
	number = {16-18},
	urldate = {2024-03-21},
	journal = {Molecular Physics},
	author = {Bowman, Joel M and Carrington, Tucker and Meyer, Hans-Dieter},
	month = aug,
	year = {2008},
	pages = {2145--2182},
}

@inproceedings{wang_computing_2015,
	address = {Kos, Greece},
	title = {Computing vibrational and ro-vibrational spectra of {CH5}+},
	url = {https://pubs.aip.org/aip/acp/article/1642/1/336-337/882914},
	doi = {10.1063/1.4906687},
	urldate = {2024-03-20},
	author = {Wang, Xiaogang and Carrington, Tucker},
	year = {2015},
	pages = {336--337},
}

@article{wang_vibrational_2008,
	title = {Vibrational energy levels of {CH5}+},
	volume = {129},
	issn = {0021-9606, 1089-7690},
	url = {https://pubs.aip.org/jcp/article/129/23/234102/349042/Vibrational-energy-levels-of-CH5},
	doi = {10.1063/1.3027825},
	
	number = {23},
	urldate = {2024-03-20},
	journal = {The Journal of Chemical Physics},
	author = {Wang, Xiao-Gang and Carrington, Tucker},
	month = dec,
	year = {2008},
	pages = {234102},
}

@article{thompson_ch_2005,
	title = {{CH} $_{\textrm{5}}$ $^{\textrm{+}}$ : {Chemistry}'s {Chameleon} {Unmasked}},
	volume = {127},
	issn = {0002-7863, 1520-5126},
	shorttitle = {{CH} $_{\textrm{5}}$ $^{\textrm{+}}$},
	url = {https://pubs.acs.org/doi/10.1021/ja0482280},
	doi = {10.1021/ja0482280},
	
	number = {13},
	urldate = {2024-03-19},
	journal = {Journal of the American Chemical Society},
	author = {Thompson, Keiran C. and Crittenden, Deborah L. and Jordan, Meredith J. T.},
	month = apr,
	year = {2005},
	pages = {4954--4958},
}

@article{schreiner_ch5_1993,
	title = {{CH}+5: {The} never-ending story or the final word?},
	volume = {99},
	issn = {0021-9606, 1089-7690},
	shorttitle = {{CH}+5},
	url = {https://pubs.aip.org/jcp/article/99/5/3716/813526/CH-5-The-never-ending-story-or-the-final-word},
	doi = {10.1063/1.466147},
	
	number = {5},
	urldate = {2024-03-19},
	journal = {The Journal of Chemical Physics},
	author = {Schreiner, Peter R. and Kim, Seung-Joon and Schaefer, Henry F. and Von Ragué Schleyer, Paul},
	month = sep,
	year = {1993},
	pages = {3716--3720},
}

@article{brown_quantum_2004,
	title = {Quantum and classical studies of vibrational motion of {CH5}+ on a global potential energy surface obtained from a novel \textit{ab initio} direct dynamics approach},
	volume = {121},
	issn = {0021-9606, 1089-7690},
	url = {https://pubs.aip.org/jcp/article/121/9/4105/186907/Quantum-and-classical-studies-of-vibrational},
	doi = {10.1063/1.1775767},
	
	number = {9},
	urldate = {2024-03-18},
	journal = {The Journal of Chemical Physics},
	author = {Brown, Alex and McCoy, Anne B. and Braams, Bastiaan J. and Jin, Zhong and Bowman, Joel M.},
	month = sep,
	year = {2004},
	pages = {4105--4116},
}

@article{wang_calculated_2016,
	title = {Calculated rotation-bending energy levels of {CH5}+ and a comparison with experiment},
	volume = {144},
	issn = {0021-9606, 1089-7690},
	url = {https://pubs.aip.org/jcp/article/144/20/204304/913110/Calculated-rotation-bending-energy-levels-of-CH-5},
	doi = {10.1063/1.4948549},
	
	number = {20},
	urldate = {2024-03-18},
	journal = {The Journal of Chemical Physics},
	author = {Wang, Xiao-Gang and Carrington, Tucker},
	month = may,
	year = {2016},
	pages = {204304},
}

@article{jin_ab_2006,
	title = {An ab {Initio} {Based} {Global} {Potential} {Energy} {Surface} {Describing} {CH} $_{\textrm{5}}$ $^{\textrm{+}}$ → {CH} $_{\textrm{3}}$ $^{\textrm{+}}$ + {H} $_{\textrm{2}}$},
	volume = {110},
	issn = {1089-5639, 1520-5215},
	url = {https://pubs.acs.org/doi/10.1021/jp053848o},
	doi = {10.1021/jp053848o},
	
	number = {4},
	urldate = {2024-03-18},
	journal = {The Journal of Physical Chemistry A},
	author = {Jin, Zhong and Braams, Bastiaan J. and Bowman, Joel M.},
	month = feb,
	year = {2006},
	pages = {1569--1574},
}

@article{white_ch_1999,
	title = {{CH} $_{\textrm{5}}$ $^{\textrm{+}}$ : {The} {Infrared} {Spectrum} {Observed}},
	volume = {284},
	issn = {0036-8075, 1095-9203},
	shorttitle = {{CH} $_{\textrm{5}}$ $^{\textrm{+}}$},
	url = {https://www.science.org/doi/10.1126/science.284.5411.135},
	doi = {10.1126/science.284.5411.135},
	
	number = {5411},
	urldate = {2024-03-16},
	journal = {Science},
	author = {White, Edmund T. and Tang, Jian and Oka, Takeshi},
	month = apr,
	year = {1999},
	pages = {135--137},
}

@article{zhang_neural_2024,
	title = {Neural canonical transformations for vibrational spectra of molecules},
	volume = {161},
	issn = {0021-9606, 1089-7690},
	url = {https://pubs.aip.org/jcp/article/161/2/024103/3302449/Neural-canonical-transformations-for-vibrational},
	doi = {10.1063/5.0209255},
	
	number = {2},
	urldate = {2024-07-11},
	journal = {The Journal of Chemical Physics},
	author = {Zhang, Qi and Wang, Rui-Si and Wang, Lei},
	month = jul,
	year = {2024},
	pages = {024103},
}

@article{brackertz_searching_2017,
	title = {Searching for new symmetry species of {CH} 5 + – {From} lines to states without a model},
	volume = {342},
	issn = {00222852},
	url = {https://linkinghub.elsevier.com/retrieve/pii/S0022285217301959},
	doi = {10.1016/j.jms.2017.08.008},
	
	urldate = {2024-06-27},
	journal = {Journal of Molecular Spectroscopy},
	author = {Brackertz, Stefan and Schlemmer, Stephan and Asvany, Oskar},
	month = dec,
	year = {2017},
	pages = {73--82},
}

@article{schmiedt_rotation-vibration_2017,
	title = {Rotation-vibration motion of extremely flexible molecules – {The} molecular superrotor},
	volume = {672},
	issn = {00092614},
	url = {https://linkinghub.elsevier.com/retrieve/pii/S0009261417300684},
	doi = {10.1016/j.cplett.2017.01.045},
	
	urldate = {2024-06-27},
	journal = {Chemical Physics Letters},
	author = {Schmiedt, Hanno and Jensen, Per and Schlemmer, Stephan},
	month = mar,
	year = {2017},
	pages = {34--46},
}

@article{schmiedt_collective_2016,
	title = {Collective {Molecular} {Superrotation}: {A} {Model} for {Extremely} {Flexible} {Molecules} {Applied} to {Protonated} {Methane}},
	volume = {117},
	copyright = {http://link.aps.org/licenses/aps-default-license},
	issn = {0031-9007, 1079-7114},
	shorttitle = {Collective {Molecular} {Superrotation}},
	url = {https://link.aps.org/doi/10.1103/PhysRevLett.117.223002},
	doi = {10.1103/PhysRevLett.117.223002},
	
	number = {22},
	urldate = {2024-06-27},
	journal = {Physical Review Letters},
	author = {Schmiedt, Hanno and Jensen, Per and Schlemmer, Stephan},
	month = nov,
	year = {2016},
	pages = {223002},
}

@article{asvany_experimental_2015,
	title = {Experimental ground-state combination differences of {CH} $_{\textrm{5}}$ $^{\textrm{+}}$},
	volume = {347},
	copyright = {http://www.sciencemag.org/about/science-licenses-journal-article-reuse},
	issn = {0036-8075, 1095-9203},
	url = {https://www.science.org/doi/10.1126/science.aaa3304},
	doi = {10.1126/science.aaa3304},
	
	number = {6228},
	urldate = {2024-06-27},
	journal = {Science},
	author = {Asvany, Oskar and Yamada, Koichi M. T. and Brünken, Sandra and Potapov, Alexey and Schlemmer, Stephan},
	month = mar,
	year = {2015},
	pages = {1346--1349},
}

@article{marx_ch_1999,
	title = {{CH} $_{\textrm{5}}$ $^{\textrm{+}}$ : {The} {Cheshire} {Cat} {Smiles}},
	volume = {284},
	issn = {0036-8075, 1095-9203},
	shorttitle = {{CH} $_{\textrm{5}}$ $^{\textrm{+}}$},
	doi = {10.1126/science.284.5411.59},
	
	number = {5411},
	urldate = {2024-03-16},
	journal = {Science},
	author = {Marx, Dominik and Parrinello, Michele},
	month = apr,
	year = {1999},
	pages = {59--61},
}

@article{huang_deuteration_2006,
	title = {Deuteration {Effects} on the {Structure} and {Infrared} {Spectrum} of {CH} $_{\textrm{5}}$ $^{\textrm{+}}$},
	volume = {128},
	issn = {0002-7863, 1520-5126},
	url = {https://pubs.acs.org/doi/10.1021/ja057514o},
	doi = {10.1021/ja057514o},
	
	number = {11},
	urldate = {2024-06-25},
	journal = {Journal of the American Chemical Society},
	author = {Huang, Xinchuan and Johnson, Lindsay M. and Bowman, Joel M. and McCoy, Anne B.},
	month = mar,
	year = {2006},
	pages = {3478--3479},
}

@article{johnson_evolution_2006,
	title = {Evolution of {Structure} in {CH} $_{\textrm{5}}$ $^{\textrm{+}}$ and {Its} {Deuterated} {Analogues}},
	volume = {110},
	issn = {1089-5639, 1520-5215},
	url = {https://pubs.acs.org/doi/10.1021/jp061675c},
	doi = {10.1021/jp061675c},
	
	number = {26},
	urldate = {2024-06-23},
	journal = {The Journal of Physical Chemistry A},
	author = {Johnson, Lindsay M. and McCoy, Anne B.},
	month = jul,
	year = {2006},
	pages = {8213--8220},
}

@article{brown_classical_2003,
	title = {Classical and quasiclassical spectral analysis of {CH5}+ using an \textit{ab initio} potential energy surface},
	volume = {119},
	issn = {0021-9606, 1089-7690},
	url = {https://pubs.aip.org/jcp/article/119/17/8790/442017/Classical-and-quasiclassical-spectral-analysis-of},
	doi = {10.1063/1.1622379},
	
	number = {17},
	urldate = {2024-06-13},
	journal = {The Journal of Chemical Physics},
	author = {Brown, Alex and Braams, Bastiaan J. and Christoffel, Kurt and Jin, Zhong and Bowman, Joel M.},
	month = nov,
	year = {2003},
	pages = {8790--8793},
}

@article{mccoy_ab_2004,
	title = {Ab {Initio} {Diffusion} {Monte} {Carlo} {Calculations} of the {Quantum} {Behavior} of {CH} $_{\textrm{5}}$ $^{\textrm{+}}$ in {Full} {Dimensionality}},
	volume = {108},
	issn = {1089-5639, 1520-5215},
	url = {https://pubs.acs.org/doi/10.1021/jp0487096},
	doi = {10.1021/jp0487096},
	
	number = {23},
	urldate = {2024-06-13},
	journal = {The Journal of Physical Chemistry A},
	author = {McCoy, Anne B. and Braams, Bastiaan J. and Brown, Alex and Huang, Xinchuan and Jin, Zhong and Bowman, Joel M.},
	month = jun,
	year = {2004},
	pages = {4991--4994},
}

@article{marx_structural_1995,
	title = {Structural quantum effects and three-centre two-electron bonding in {CH}+5},
	volume = {375},
	copyright = {http://www.springer.com/tdm},
	issn = {0028-0836, 1476-4687},
	url = {https://www.nature.com/articles/375216a0},
	doi = {10.1038/375216a0},
	
	number = {6528},
	urldate = {2025-04-10},
	journal = {Nature},
	author = {Marx, Dominik and Parrinello, Michele},
	month = may,
	year = {1995},
	pages = {216--218},
}

@article{hinkle_characterizing_2008,
	title = {Characterizing {Excited} {States} of {CH}$_{\textrm{5}}$$^{\textrm{+}}$ with {Diffusion} {Monte} {Carlo}},
	volume = {112},
	issn = {1089-5639, 1520-5215},
	url = {https://pubs.acs.org/doi/10.1021/jp709828v},
	doi = {10.1021/jp709828v},
	
	number = {10},
	urldate = {2025-04-10},
	journal = {The Journal of Physical Chemistry A},
	author = {Hinkle, Charlotte E. and McCoy, Anne B.},
	month = mar,
	year = {2008},
	pages = {2058--2064},
}

@article{bunker_theoretical_2004,
	title = {A theoretical study of the millimeterwave spectrum of {CH5}+},
	volume = {695-696},
	copyright = {https://www.elsevier.com/tdm/userlicense/1.0/},
	issn = {00222860},
	url = {https://linkinghub.elsevier.com/retrieve/pii/S0022286003008603},
	doi = {10.1016/j.molstruc.2003.12.020},
	
	urldate = {2025-04-15},
	journal = {Journal of Molecular Structure},
	author = {Bunker, P.R. and Ostojić, B. and Yurchenko, S.},
	month = jun,
	year = {2004},
	pages = {253--261},
}

@article{simko_quantum-chemical_2023,
	title = {Quantum-{Chemical} and {Quantum}-{Graph} {Models} of the {Dynamical} {Structure} of {CH}$_{\textrm{5}}$$^{\textrm{+}}$},
	volume = {19},
	copyright = {https://doi.org/10.15223/policy-029},
	issn = {1549-9618, 1549-9626},
	url = {https://pubs.acs.org/doi/10.1021/acs.jctc.2c00991},
	doi = {10.1021/acs.jctc.2c00991},
	
	number = {1},
	urldate = {2025-04-19},
	journal = {Journal of Chemical Theory and Computation},
	author = {Simkó, Irén and Fábri, Csaba and Császár, Attila G.},
	month = jan,
	year = {2023},
	pages = {42--50},
}

@article{fabri_use_2017,
	title = {On the use of nonrigid-molecular symmetry in nuclear motion computations employing a discrete variable representation: {A} case study of the bending energy levels of {CH5}+},
	volume = {147},
	issn = {0021-9606, 1089-7690},
	shorttitle = {On the use of nonrigid-molecular symmetry in nuclear motion computations employing a discrete variable representation},
	url = {https://pubs.aip.org/jcp/article/147/13/134101/195717/On-the-use-of-nonrigid-molecular-symmetry-in},
	doi = {10.1063/1.4990297},
	
	number = {13},
	urldate = {2025-04-19},
	journal = {The Journal of Chemical Physics},
	author = {Fábri, Csaba and Quack, Martin and Császár, Attila G.},
	month = oct,
	year = {2017},
	pages = {134101},
}

@article{wodraszka_ch5_2015,
	title = {{CH}$_{\textrm{5}}$$^{\textrm{+}}$ : {Symmetry} and the {Entangled} {Rovibrational} {Quantum} {States} of a {Fluxional} {Molecule}},
	volume = {6},
	issn = {1948-7185, 1948-7185},
	shorttitle = {{CH}$_{\textrm{5}}$$^{\textrm{+}}$},
	url = {https://pubs.acs.org/doi/10.1021/acs.jpclett.5b01869},
	doi = {10.1021/acs.jpclett.5b01869},
	
	number = {21},
	urldate = {2025-04-19},
	journal = {The Journal of Physical Chemistry Letters},
	author = {Wodraszka, Robert and Manthe, Uwe},
	month = nov,
	year = {2015},
	pages = {4229--4232},
}

@article{hinkle_theoretical_2009,
	title = {Theoretical {Investigations} of {Mode} {Mixing} in {Vibrationally} {Excited} {States} of {CH}$_{\textrm{5}}$$^{\textrm{+}}$},
	volume = {113},
	issn = {1089-5639, 1520-5215},
	url = {https://pubs.acs.org/doi/10.1021/jp8112733},
	doi = {10.1021/jp8112733},
	
	number = {16},
	urldate = {2025-04-28},
	journal = {The Journal of Physical Chemistry A},
	author = {Hinkle, Charlotte E. and McCoy, Anne B.},
	month = apr,
	year = {2009},
	pages = {4587--4597},
}

@article{li_neural_2020,
	title = {Neural {Canonical} {Transformation} with {Symplectic} {Flows}},
	volume = {10},
	issn = {2160-3308},
	url = {http://arxiv.org/abs/1910.00024},
	doi = {10.1103/PhysRevX.10.021020},
	
	number = {2},
	urldate = {2025-05-11},
	journal = {Physical Review X},
	author = {Li, Shuo-Hui and Dong, Chen-Xiao and Zhang, Linfeng and Wang, Lei},
	month = apr,
	year = {2020},
	note = {arXiv:1910.00024 [cond-mat]},
	pages = {021020},
}

@article{rawlinson_exactly_2021,
	title = {Exactly solvable {1D} model explains the low-energy vibrational level structure of protonated methane},
	volume = {57},
	issn = {1359-7345, 1364-548X},
	url = {https://xlink.rsc.org/?DOI=D1CC01214B},
	doi = {10.1039/D1CC01214B},
	
	number = {39},
	urldate = {2025-05-19},
	journal = {Chemical Communications},
	author = {Rawlinson, Jonathan I. and Fábri, Csaba and Császár, Attila G.},
	year = {2021},
	pages = {4827--4830},
}

@article{fabri_vibrational_2018,
	title = {Vibrational quantum graphs and their application to the quantum dynamics of {CH}$_{\textrm{5}}$$^{\textrm{+}}$},
	volume = {20},
	issn = {1463-9076, 1463-9084},
	url = {https://xlink.rsc.org/?DOI=C8CP03019G},
	doi = {10.1039/C8CP03019G},
	
	number = {25},
	urldate = {2025-05-19},
	journal = {Physical Chemistry Chemical Physics},
	author = {Fábri, Csaba and Császár, Attila G.},
	year = {2018},
	pages = {16913--16917},
}

@article{shanavas_rasheeda_high-dimensional_2022,
	title = {High-dimensional neural network potentials for accurate vibrational frequencies: the formic acid dimer benchmark},
	volume = {24},
	issn = {1463-9076, 1463-9084},
	shorttitle = {High-dimensional neural network potentials for accurate vibrational frequencies},
	url = {https://xlink.rsc.org/?DOI=D2CP03893E},
	doi = {10.1039/D2CP03893E},
	
	number = {48},
	urldate = {2025-05-20},
	journal = {Physical Chemistry Chemical Physics},
	author = {Shanavas Rasheeda, Dilshana and Martín Santa Daría, Alberto and Schröder, Benjamin and Mátyus, Edit and Behler, Jörg},
	year = {2022},
	pages = {29381--29392},
}

@article{ren_machine_2021,
	title = {A machine learning vibrational spectroscopy protocol for spectrum prediction and spectrum-based structure recognition},
	volume = {1},
	issn = {26673258},
	url = {https://linkinghub.elsevier.com/retrieve/pii/S2667325821000972},
	doi = {10.1016/j.fmre.2021.05.005},
	
	number = {4},
	urldate = {2025-05-20},
	journal = {Fundamental Research},
	author = {Ren, Hao and Li, Hao and Zhang, Qian and Liang, Lijun and Guo, Wenyue and Huang, Fang and Luo, Yi and Jiang, Jun},
	month = jul,
	year = {2021},
	pages = {488--494},
}

@incollection{manzhos_machine_2023,
	title = {Machine learning for vibrational spectroscopy},
	copyright = {https://www.elsevier.com/tdm/userlicense/1.0/},
	isbn = {978-0-323-90049-2},
	url = {https://linkinghub.elsevier.com/retrieve/pii/B9780323900492000275},
	
	urldate = {2025-05-20},
	booktitle = {Quantum {Chemistry} in the {Age} of {Machine} {Learning}},
	publisher = {Elsevier},
	author = {Manzhos, Sergei and Ihara, Manabu and Carrington, Tucker},
	year = {2023},
	doi = {10.1016/B978-0-323-90049-2.00027-5},
	pages = {355--390},
}

@article{manzhos_neural_2021,
	title = {Neural {Network} {Potential} {Energy} {Surfaces} for {Small} {Molecules} and {Reactions}},
	volume = {121},
	copyright = {https://doi.org/10.15223/policy-029},
	issn = {0009-2665, 1520-6890},
	url = {https://pubs.acs.org/doi/10.1021/acs.chemrev.0c00665},
	doi = {10.1021/acs.chemrev.0c00665},
	
	number = {16},
	urldate = {2025-05-20},
	journal = {Chemical Reviews},
	author = {Manzhos, Sergei and Carrington, Tucker},
	month = aug,
	year = {2021},
	pages = {10187--10217},
}

@article{han_concise_2022,
	title = {A {Concise} {Review} on {Recent} {Developments} of {Machine} {Learning} for the {Prediction} of {Vibrational} {Spectra}},
	volume = {126},
	copyright = {https://doi.org/10.15223/policy-029},
	issn = {1089-5639, 1520-5215},
	url = {https://pubs.acs.org/doi/10.1021/acs.jpca.1c10417},
	doi = {10.1021/acs.jpca.1c10417},
	
	number = {6},
	urldate = {2025-05-20},
	journal = {The Journal of Physical Chemistry A},
	author = {Han, Ruocheng and Ketkaew, Rangsiman and Luber, Sandra},
	month = feb,
	year = {2022},
	pages = {801--812},
}

@article{saleh_computing_2025,
	title = {Computing {Excited} {States} of {Molecules} {Using} {Normalizing} {Flows}},
	copyright = {https://creativecommons.org/licenses/by/4.0/},
	issn = {1549-9618, 1549-9626},
	url = {https://pubs.acs.org/doi/10.1021/acs.jctc.5c00590},
	doi = {10.1021/acs.jctc.5c00590},
	
	urldate = {2025-05-20},
	journal = {Journal of Chemical Theory and Computation},
	author = {Saleh, Yahya and Fern{\'a}ndez Corral, {\'A}lvaro and Vogt, Emil and Iske, Armin and K{\"u}pper, Jochen and Yachmenev, Andrey},
	month = may,
	year = {2025},
	pages = {acs.jctc.5c00590},
}

@article{ishii_development_2022,
	title = {Development of anharmonic vibrational structure theory using backflow transformation},
	volume = {787},
	issn = {00092614},
	url = {https://linkinghub.elsevier.com/retrieve/pii/S0009261421009465},
	doi = {10.1016/j.cplett.2021.139263},
	
	urldate = {2025-05-20},
	journal = {Chemical Physics Letters},
	author = {Ishii, Kiriko and Shimazaki, Tomomi and Tachikawa, Masanori and Kita, Yukiumi},
	month = jan,
	year = {2022},
	pages = {139263},
}

@article{han_ai-powered_2025,
	title = {{AI}-powered exploration of molecular vibrations, phonons, and spectroscopy},
	volume = {4},
	issn = {2635-098X},
	url = {https://xlink.rsc.org/?DOI=D4DD00353E},
	doi = {10.1039/D4DD00353E},
	
	number = {3},
	urldate = {2025-05-20},
	journal = {Digital Discovery},
	author = {Han, Bowen and Okabe, Ryotaro and Chotrattanapituk, Abhijatmedhi and Cheng, Mouyang and Li, Mingda and Cheng, Yongqiang},
	year = {2025},
	pages = {584--624},
}

@article{hao_xie_ab-initio_2022,
	title = {Ab-{Initio} {Study} of {Interacting} {Fermions} at {Finite} {Temperature} with {Neural} {Canonical} {Transformation}},
	volume = {1},
	copyright = {https://creativecommons.org/licenses/by/4.0/},
	issn = {2790-2048, 2790-203X},
	url = {https://global-sci.com/article/87606/ab-initio-study-of-interacting-fermions-at-finite-temperature-with-neural-canonical-transformation},
	doi = {10.4208/jml.220113},
	number = {1},
	urldate = {2025-05-21},
	journal = {Journal of Machine Learning},
	author = {Xie, Hao and Zhang, Linfeng and Wang, Lei},
	month = jan,
	year = {2022},
	pages = {38--59},
}

@misc{zhang_neural_2024-1,
	title = {Neural {Canonical} {Transformations} for {Quantum} {Anharmonic} {Solids} of {Lithium}},
	copyright = {arXiv.org perpetual, non-exclusive license},
	url = {https://arxiv.org/abs/2412.12451},
	doi = {10.48550/ARXIV.2412.12451},
	urldate = {2025-05-21},
	publisher = {arXiv},
	author = {Zhang, Qi and Wang, Xiaoyang and Shi, Rong and Ren, Xinguo and Wang, Han and Wang, Lei},
	year = {2024},
	note = {Version Number: 2},
}

@article{xie_m_2023,
	title = {\$m{\textasciicircum}*\$ of two-dimensional electron gas: {A} neural canonical transformation study},
	volume = {14},
	issn = {2542-4653},
	shorttitle = {\$m{\textasciicircum}*\$ of two-dimensional electron gas},
	url = {https://scipost.org/10.21468/SciPostPhys.14.6.154},
	doi = {10.21468/SciPostPhys.14.6.154},
	number = {6},
	urldate = {2025-05-21},
	journal = {SciPost Physics},
	author = {Xie, Hao and Zhang, Linfeng and Wang, Lei},
	month = jun,
	year = {2023},
	pages = {154},
}

@misc{papamakarios_normalizing_2021,
	title = {Normalizing {Flows} for {Probabilistic} {Modeling} and {Inference}},
	url = {http://arxiv.org/abs/1912.02762},
	doi = {10.48550/arXiv.1912.02762},
	urldate = {2025-05-25},
	publisher = {arXiv},
	author = {Papamakarios, George and Nalisnick, Eric and Rezende, Danilo Jimenez and Mohamed, Shakir and Lakshminarayanan, Balaji},
	month = apr,
	year = {2021},
	note = {arXiv:1912.02762 [stat]},
}

@article{kobyzev_normalizing_2021,
	title = {Normalizing {Flows}: {An} {Introduction} and {Review} of {Current} {Methods}},
	volume = {43},
	copyright = {https://ieeexplore.ieee.org/Xplorehelp/downloads/license-information/IEEE.html},
	issn = {0162-8828, 2160-9292, 1939-3539},
	shorttitle = {Normalizing {Flows}},
	url = {https://ieeexplore.ieee.org/document/9089305/},
	doi = {10.1109/TPAMI.2020.2992934},
	number = {11},
	urldate = {2025-05-25},
	journal = {IEEE Transactions on Pattern Analysis and Machine Intelligence},
	author = {Kobyzev, Ivan and Prince, Simon J.D. and Brubaker, Marcus A.},
	month = nov,
	year = {2021},
	pages = {3964--3979},
}

@misc{dinh_density_2017,
	title = {Density estimation using {Real} {NVP}},
	url = {http://arxiv.org/abs/1605.08803},
	doi = {10.48550/arXiv.1605.08803},
	urldate = {2025-05-25},
	publisher = {arXiv},
	author = {Dinh, Laurent and Sohl-Dickstein, Jascha and Bengio, Samy},
	month = feb,
	year = {2017},
	note = {arXiv:1605.08803 [cs]},
}

@incollection{courant_methods_2009,
	address = {Weinheim},
	title = {Methods of mathematical physics. {Vol}.1},
	isbn = {978-0-471-50447-4},
	publisher = {Wiley-VCH},
	author = {Courant, Richard and Hilbert, David},
	year = {2009},
	note = {Num Pages: 456},
}

@book{becca_quantum_2017,
	address = {Cambridge New York},
	title = {Quantum {Monte} {Carlo} approaches for correlated systems},
	isbn = {978-1-107-12993-1},
	publisher = {Cambridge university pres},
	author = {Becca, Federico and Sorella, Sandro},
	year = {2017},
}

@misc{kingma_adam_2017,
	title = {Adam: {A} {Method} for {Stochastic} {Optimization}},
	shorttitle = {Adam},
	url = {http://arxiv.org/abs/1412.6980},
	doi = {10.48550/arXiv.1412.6980},
	urldate = {2025-05-25},
	publisher = {arXiv},
	author = {Kingma, Diederik P. and Ba, Jimmy},
	month = jan,
	year = {2014},
	note = {arXiv:1412.6980 [cs]},
}

@inproceedings{tal1952and,
  title={and AK Lyubimova},
  author={Tal'roze, V},
  booktitle={Doklady Akad. Nauk SSSR},
  volume={86},
  pages={909},
  year={1952}
}

@article{rawlinson_quantum_2019,
	title = {Quantum graph model for rovibrational states of protonated methane},
	volume = {151},
	issn = {0021-9606, 1089-7690},
	url = {https://pubs.aip.org/jcp/article/151/16/164303/1065365/Quantum-graph-model-for-rovibrational-states-of},
	doi = {10.1063/1.5125986},
	number = {16},
	urldate = {2025-05-28},
	journal = {The Journal of Chemical Physics},
	author = {Rawlinson, J. I.},
	month = oct,
	year = {2019},
	pages = {164303},
}

@article{olah1995electrophilic,
  title={Electrophilic substitution of methane revisited},
  author={Olah, George A and Hartz, Nikolai and Rasul, Golam and Prakash, GK Surya},
  journal={Journal of the American Chemical Society},
  volume={117},
  number={4},
  pages={1336--1343},
  year={1995},
  publisher={ACS Publications}
}

@article{olah1997kekule,
  title={From Kekul{\'e}'s tetravalent methane to five-, six-, and seven-coordinate protonated methanes},
  author={Olah, George A and Rasul, Golam},
  journal={Accounts of chemical research},
  volume={30},
  number={6},
  pages={245--250},
  year={1997},
  publisher={ACS Publications}
}

@article{he03000u,
  author={Herbst, E. and Green, S. and Thaddeus, P. and Klemperer, W.},
  title={Indirect observation of unobservable interstellar molecules},
  year={1977},
  journal={Astrophysical Journal},
  volume={215},
  pages={503--510},
  doi={10.1086/155381},
}

@article{talbi1992quantum,
  title={Quantum chemical calculations for a better understanding of the mechanism of CH3 (+)+ H2 radiative association},
  author={Talbi, D and Saxon, RP},
  journal={Astronomy and Astrophysics (ISSN 0004-6361), vol. 261, no. 2, p. 671-676.},
  volume={261},
  pages={671--676},
  year={1992}
}

@article{momma_vesta_2011,
	title = {\textit{{VESTA} 3} for three-dimensional visualization of crystal, volumetric and morphology data},
	volume = {44},
	issn = {0021-8898},
	url = {https://journals.iucr.org/paper?S0021889811038970},
	doi = {10.1107/S0021889811038970},
	number = {6},
	urldate = {2025-06-08},
	journal = {Journal of Applied Crystallography},
	author = {Momma, Koichi and Izumi, Fujio},
	month = dec,
	year = {2011},
	pages = {1272--1276},
}

@misc{zhang2025quantumanharmoniceffectshydrogenbond,
      title={Quantum Anharmonic Effects in Hydrogen-Bond Symmetrization of High-Pressure Ice}, 
      author={Qi Zhang and Lei Wang},
      year={2025},
      eprint={2507.01452},
      archivePrefix={arXiv},
      primaryClass={cond-mat.mtrl-sci},
      url={https://arxiv.org/abs/2507.01452}, 
}

@misc{li2025deepvariationalfreeenergy,
      title={Deep Variational Free Energy Calculation of Hydrogen Hugoniot}, 
      author={Zihang Li and Hao Xie and Xinyang Dong and Lei Wang},
      year={2025},
      eprint={2507.18540},
      archivePrefix={arXiv},
      primaryClass={cond-mat.str-el},
      url={https://arxiv.org/abs/2507.18540}, 
}

@article{wirnsberger_targeted_2020,
	title = {Targeted free energy estimation via learned mappings},
	volume = {153},
	issn = {0021-9606, 1089-7690},
	url = {https://pubs.aip.org/jcp/article/153/14/144112/316574/Targeted-free-energy-estimation-via-learned},
	doi = {10.1063/5.0018903},
	number = {14},
	urldate = {2025-09-24},
	journal = {The Journal of Chemical Physics},
	author = {Wirnsberger, Peter and Ballard, Andrew J. and Papamakarios, George and Abercrombie, Stuart and Racanière, Sébastien and Pritzel, Alexander and Jimenez Rezende, Danilo and Blundell, Charles},
	month = oct,
	year = {2020},
	pages = {144112},
}

@article{gross_rayleigh-ritz_1988,
	title = {Rayleigh-Ritz variational principle for ensembles of fractionally occupied states},
	volume = {37},
	rights = {http://link.aps.org/licenses/aps-default-license},
	issn = {0556-2791},
	url = {https://link.aps.org/doi/10.1103/PhysRevA.37.2805},
	doi = {10.1103/PhysRevA.37.2805},
	pages = {2805--2808},
	number = {8},
	journal = {Physical Review A},
	shortjournal = {Phys. Rev. A},
	author = {Gross, E. K. U. and Oliveira, L. N. and Kohn, W.},
	year = {1988},
	urldate = {2024-07-17},
	date = {1988-04-01},
	langid = {english},
}

@article{eckart_studies_1935,
	title = {Some Studies Concerning Rotating Axes and Polyatomic Molecules},
	volume = {47},
	issn = {0031-899X},
	url = {https://link.aps.org/doi/10.1103/PhysRev.47.552},
	doi = {10.1103/PhysRev.47.552},
	pages = {552--558},
	number = {7},
	journal = {Physical Review},
	shortjournal = {Phys. Rev.},
	author = {Eckart, Carl},
	date = {1935-04-01},
	langid = {english},
}

@article{williams_simple_1992,
	title = {Simple statistical gradient-following algorithms for connectionist reinforcement learning},
	volume = {8},
	copyright = {http://www.springer.com/tdm},
	issn = {0885-6125, 1573-0565},
	url = {http://link.springer.com/10.1007/BF00992696},
	doi = {10.1007/BF00992696},
	number = {3-4},
	urldate = {2026-04-11},
	journal = {Machine Learning},
	author = {Williams, Ronald J.},
	month = may,
	year = {1992},
	pages = {229--256},
}

@article{noe_boltzmann_2019,
	title = {Boltzmann generators: {Sampling} equilibrium states of many-body systems with deep learning},
	volume = {365},
	issn = {0036-8075, 1095-9203},
	shorttitle = {Boltzmann generators},
	url = {https://www.science.org/doi/10.1126/science.aaw1147},
	doi = {10.1126/science.aaw1147},
	number = {6457},
	urldate = {2026-04-14},
	journal = {Science},
	author = {Noé, Frank and Olsson, Simon and Köhler, Jonas and Wu, Hao},
	month = sep,
	year = {2019},
	pages = {eaaw1147},
}

@misc{jin_v2rho-fno_2026,
	title = {{V2Rho}-{FNO}: {Fourier} {Neural} {Operator} for {Electronic} {Density} {Prediction}},
	shorttitle = {{V2Rho}-{FNO}},
	url = {http://arxiv.org/abs/2603.15669},
	doi = {10.48550/arXiv.2603.15669},
	urldate = {2026-04-14},
	publisher = {arXiv},
	author = {Jin, Yingdi and Qin, Xinming and Liu, Ruichen and Liu, Jie and Li, Zhenyu and Yang, Jinlong},
	month = mar,
	year = {2026},
	note = {arXiv:2603.15669 [physics]},
}

@misc{shah_fourier_2026,
	title = {Fourier {Neural} {Operators} for {Learning} {Dynamics} in {Quantum} {Spin} {Systems}},
	url = {http://arxiv.org/abs/2409.03302},
	doi = {10.48550/arXiv.2409.03302},
	urldate = {2026-04-14},
	publisher = {arXiv},
	author = {Shah, Freya and Patti, Taylor L. and Berner, Julius and Tolooshams, Bahareh and Kossaifi, Jean and Anandkumar, Anima},
	month = jan,
	year = {2026},
	note = {arXiv:2409.03302 [quant-ph]},
}

@article{mizera_scattering_2023,
	title = {Scattering with neural operators},
	volume = {108},
	issn = {2470-0010, 2470-0029},
	url = {https://link.aps.org/doi/10.1103/PhysRevD.108.L101701},
	doi = {10.1103/PhysRevD.108.L101701},
	number = {10},
	urldate = {2026-04-14},
	journal = {Physical Review D},
	author = {Mizera, Sebastian},
	month = nov,
	year = {2023},
	pages = {L101701},
}

@article{kovachki2023neural,
  title={Neural operator: Learning maps between function spaces with applications to pdes},
  author={Kovachki, Nikola and Li, Zongyi and Liu, Burigede and Azizzadenesheli, Kamyar and Bhattacharya, Kaushik and Stuart, Andrew and Anandkumar, Anima},
  journal={Journal of Machine Learning Research},
  volume={24},
  number={89},
  pages={1--97},
  year={2023}
}

@misc{li_fourier_2021,
	title = {Fourier {Neural} {Operator} for {Parametric} {Partial} {Differential} {Equations}},
	url = {http://arxiv.org/abs/2010.08895},
	doi = {10.48550/arXiv.2010.08895},
	urldate = {2026-04-14},
	publisher = {arXiv},
	author = {Li, Zongyi and Kovachki, Nikola and Azizzadenesheli, Kamyar and Liu, Burigede and Bhattacharya, Kaushik and Stuart, Andrew and Anandkumar, Anima},
	month = may,
	year = {2021},
	note = {arXiv:2010.08895 [cs]},
}

@article{10.1063/5.0285954,
    author = {Vogt, Emil and Fern{\'a}ndez Corral, {\'A}lvaro and Saleh, Yahya and Yachmenev, Andrey},
    title = {Transferability and interpretability of vibrational normalizing-flow coordinates},
    journal = {The Journal of Chemical Physics},
    volume = {163},
    number = {15},
    pages = {154106},
    year = {2025},
    month = {10},
    issn = {0021-9606},
    doi = {10.1063/5.0285954},
    url = {https://doi.org/10.1063/5.0285954},
}

@misc{saleh_convergence_2026,
      title={Convergence theory for Hermite approximations under adaptive coordinate transformations},
      author={Yahya Saleh},
      year={2026},
      eprint={2604.16975},
      archivePrefix={arXiv},
      primaryClass={math.NA},
      url={https://arxiv.org/abs/2604.16975},
}
\clearpage

\onecolumngrid

\setcounter{algorithm}{0}
\renewcommand{\thealgorithm}{\arabic{algorithm}}
\setcounter{table}{0}
\renewcommand{\thetable}{S\arabic{table}}
\setcounter{figure}{0}
\renewcommand{\thefigure}{S\arabic{figure}}
\setcounter{equation}{0}
\renewcommand{\theequation}{S\arabic{equation}}

\appendix*

\section{Supplementary Information}
\label{append:supplement}

\subsection{Eckart-Consistent MCMC Sampling}
\label{subsec:eckart}

In the main text, the MCMC sampling is performed in the $J{=}0$ pure vibrational subspace. This is achieved by constructing proposals that live entirely within the Eckart frame~\cite{eckart_studies_1935}, a body-fixed coordinate system that decouples internal vibrational motion from overall translation and rotation.

Consider an $N$-atom molecule with masses $\{m_i\}_{i=1}^N$ and lab-frame positions $\{\vec{r}_i \in \mathbb{R}^3\}$. Fix a reference configuration $\{\vec{r}_i^{\,0}\}$ (typically an equilibrium geometry). Define the center-of-mass (COM) position $\vec{R}_{\mathrm{cm}} = M^{-1}\sum_i m_i \vec{r}_i$ with total mass $M = \sum_i m_i$, and the COM-centered coordinates
\begin{equation}
    \vec{x}_i = \vec{r}_i - \vec{R}_{\mathrm{cm}}, \qquad \vec{q}_i = \vec{r}_i^{\,0} - \vec{R}_{\mathrm{cm}}^{\,0},
    \label{eq:com_centered}
\end{equation}
where $\vec{x}_i$ denotes the COM-centered position of atom $i$ in the current configuration and $\vec{q}_i$ denotes that in the reference.

The Eckart frame is obtained by finding the rotation $R_*$ that best aligns the current configuration to the reference in the mass-weighted metric:
\begin{equation}
    R_* = \arg\min_{R} \sum_{i=1}^N m_i \|\vec{x}_i R - \vec{q}_i\|^2.
    \label{eq:eckart_rotation}
\end{equation}
The Eckart-fixed coordinates are then $\vec{y}_i = \vec{x}_i R_*$, which satisfy the Eckart rotational condition $\sum_i m_i\, \vec{q}_i \times \vec{y}_i = \vec{0}$. In practice, the optimal rotation is computed via the singular value decomposition of the $3 \times 3$ mass-weighted correlation matrix $C = \sum_i m_i\, \vec{x}_i^{\,\mathsf{T}} \vec{q}_i$. Writing $C = U \Sigma V^{\mathsf{T}}$, the optimal rotation is $R_* = U V^{\mathsf{T}}$.
We note that differentiating through the SVD is required for the kinetic energy evaluation, which involves second derivatives of the wavefunction with respect to the coordinates $\vec{x}$. However, the SVD does not enter the parameter gradient of the loss function, because the local energies are treated as constants (i.e., their dependence on $\theta$ through $\vec{x}$ is not differentiated) in the REINFORCE-type estimator of \Eq{eq:loss_gradient}.
Since $C$ is a $3 \times 3$ matrix, it is generically full-rank for a non-planar molecule such as CH$_5^+$, whose mass-weighted correlation matrix has three well-separated singular values. Near-degenerate singular values would require collinear or planar instantaneous geometries of the atoms. In our calculation, the MCMC sampler draws molecular configurations on the physical PES, and such collinear or planar geometries are strongly disfavored by the PES itself: they correspond to very high-energy regions that the sampler essentially never visits. In practice, we observed no numerical instability arising from the SVD during training.

Eckart-consistent MCMC proposal: To sample the wavefunction in the Eckart gauge, we construct proposals that contain no net translational or rotational displacement. Working in stacked mass-weighted coordinates $\delta \vec{Q} \in \mathbb{R}^{3N}$ with atom-wise blocks $\delta \vec{Q}_i = \sqrt{m_i}\,\delta \vec{y}_i$, the rigid-body directions at the reference are
\begin{align}
    \delta \vec{Q}_i^{(T,a)} &= \sqrt{m_i}\,\vec{e}_a, \label{eq:trans_mode} \\
    \delta \vec{Q}_i^{(R,a)} &= \sqrt{m_i}\,(\vec{e}_a \times \vec{q}_i), \label{eq:rot_mode}
\end{align}
for $a = x, y, z$, where $\vec{e}_a$ are the Cartesian unit vectors. Let $B \in \mathbb{R}^{3N \times k}$ ($k \leq 6$) have orthonormal columns spanning these rigid-body directions. The projector onto the internal (vibrational) subspace is
\begin{equation}
    P_{\mathrm{int}} = I - B B^{\mathsf{T}}.
    \label{eq:proj_int}
\end{equation}

The proposal proceeds as follows: (i) draw an isotropic Gaussian displacement $\delta Q \sim \mathcal{N}(0, \sigma^2 I_{3N})$; (ii) project out rigid-body components, $\delta Q_{\mathrm{int}} = P_{\mathrm{int}}\,\delta Q$; (iii) convert to Cartesian displacements, $\delta \vec{y}_i = \delta \vec{Q}_{\mathrm{int},i} / \sqrt{m_i}$; (iv) propose $\vec{y}'_i = \vec{y}_i + \delta \vec{y}_i$.

The projection guarantees $\sum_i m_i\, \delta \vec{y}_i = \vec{0}$ (no COM drift) and $\sum_i m_i\, \vec{q}_i \times \delta \vec{y}_i = \vec{0}$ (no infinitesimal rotation), so the proposal remains in the Eckart gauge. Moreover, because $P_{\mathrm{int}}$ is a symmetric projector and the base draw is isotropic, the projected increment has an even distribution, yielding a symmetric proposal $q(\vec{y}' \mid \vec{y}) = q(\vec{y} \mid \vec{y}')$. The Metropolis--Hastings acceptance ratio thus reduces to the standard Metropolis form:
\begin{equation}
    \alpha(\vec{y} \to \vec{y}') = \min\!\left(1, \frac{|\Psi(\vec{y}')|^2}{|\Psi(\vec{y})|^2}\right).
    \label{eq:mh_acceptance}
\end{equation}

\subsection{Computational Cost and Scaling with Number of States}
\label{subsec:comp_cost}

The two production calculations reported in the main text were performed on multi-GPU compute nodes with the batch distributed across devices via \texttt{jax.pmap}: the ground-state run ($N_o=1$) used $4\times$ NVIDIA RTX~5090 GPUs and converged in approximately $3$~hours ($20{,}000$ iterations); the joint $32$-state excited-state run ($N_o=32$) used $4\times$ NVIDIA A100~$80$\,GB GPUs and converged in approximately $6$~hours $14$~minutes ($10{,}000$ iterations).

To quantify how the per-iteration training cost scales with the number of simultaneously optimized states $N_o$, we performed six single-GPU measurement runs on an NVIDIA V100-SXM2 $32$\,GB GPU (16 CPU cores, $32$\,GB host RAM), varying $N_o \in \{1,2,4,8,16,32\}$ while holding all other hyperparameters fixed at the excited-state production values (RNVP depth~$16$, MLP width~$32$, per-state batch $1200$, $100$ MCMC steps per iteration, Adam learning rate $5\times 10^{-5}$, clip factor $5.0$). Each run performed $100$ training iterations. The first iteration is dominated by JAX just-in-time (JIT) compilation (${\sim}10$~s, only weakly dependent on $N_o$); we therefore report the mean per-iteration wall-clock over iterations $11$--$100$ in Table~\ref{tab:scaling_num_orb}.

\begin{table}[htbp]
    \ra{1.1}
    \centering
    \caption{Per-iteration wall-clock of the joint $N_o$-state NCT training on a single NVIDIA V100-SXM2 $32$\,GB GPU. All hyperparameters match the excited-state production run reported in the main text (RNVP depth~$16$, MLP width~$32$, per-state batch $1200$, $100$ MCMC steps); only $N_o$ varies. ``Mean $t_{\mathrm{iter}}$'' is computed over iterations $11$--$100$ to exclude JIT compilation; ``$100$-iter wall-clock'' is the total measured time for the full $100$-iteration run (including the ${\sim}10$~s JIT compile).}
    \label{tab:scaling_num_orb}
    \begin{tabular}{@{}cccc@{}}
    \toprule
    $N_o$ & mean $t_{\mathrm{iter}}$ (s) & relative to $N_o=1$ & $100$-iter wall-clock (s) \\
    \midrule
    $1$  & $0.56$ & $1.00\times$  & $65$  \\
    $2$  & $0.78$ & $1.39\times$  & $87$  \\
    $4$  & $1.10$ & $1.96\times$  & $119$ \\
    $8$  & $1.82$ & $3.25\times$  & $191$ \\
    $16$ & $3.16$ & $5.64\times$  & $327$ \\
    $32$ & $5.98$ & $10.68\times$ & $607$ \\
    \bottomrule
    \end{tabular}
\end{table}

Fitting the measured per-iteration wall-clock to $t(N_o) = t_0 + \tau N_o$ by least squares yields $t_0 \approx 0.41$~s and $\tau \approx 0.17$~s/state, verifying the linear scaling in $N_o$.

\begin{table}[htbp]
    \ra{1.1}
    \centering
    \caption{Vibrational energy levels of CH$_5^+$ from our NCT calculation, compared against two independent high-precision references on the same JBB PES~\cite{jin_ab_2006}. ``(Var)'' denotes the converged full-dimensional (12D, Basis~IV) variational results of Wang \& Carrington~\cite{wang_vibrational_2008}; ``(DMC)'' denotes the fixed-node diffusion Monte Carlo results of Hinkle \& McCoy~\cite{hinkle_characterizing_2008}. The alignment is a best numerical match between the two ascending sequences; reference levels with no close NCT counterpart are shown on rows with the NCT columns blank, and vice versa. The table continues from the first group of three columns into the second group, still in ascending order. All energies are in cm$^{-1}$, reported relative to the zero-point energy; the zero-point energy is omitted. Our NCT calculation does not enforce the hydrogen-permutation symmetry, so we do not attempt to match states by irreducible representation.}
    \label{tab:full_energy}
\begin{tabular}{@{}cccccc@{}}
\toprule
\textbf{\#} & \textbf{NCT Energy} & \textbf{Reference} & \textbf{\#} & \textbf{NCT Energy} & \textbf{Reference} \\
\midrule
1  & 9.7(0.1)  & 10.4~(Var)   & 14 & 459.8(0.3)  &               \\
   &           & 21.7~(Var)   & 15 & 468.6(0.3)  &               \\
2  & 34.3(0.2) &              & 16 & 473.9(0.3)  &               \\
3  & 38.2(0.2) & 39.3~(Var)   & 17 & 476.1(0.3)  &               \\
   &           & 39.8~(Var)   & 18 & 606.2(0.3)  &               \\
   &           & 44(5)~(DMC)  & 19 & 625.3(0.2)  &               \\
   &           & 47.3~(Var)   & 20 & 991.1(0.4)  & 990(5)~(DMC)  \\
   &           & 52.3~(Var)   & 21 & 995.1(0.4)  &               \\
4  & 71.9(0.2) & 85.6~(Var)   & 22 & 1040.4(0.3) &               \\
5  & 79.1(0.2) & 89.4~(Var)   & 23 & 1042.0(0.4) &               \\
6  & 97.2(0.2) & 96.2~(Var)   & 24 & 1237.2(0.3) & 1128(5)~(DMC) \\
   &           & 106.5~(Var)  & 25 & 1394.6(0.4) &               \\
   &           & 137.1~(Var)  & 26 & 1404.2(0.3) &               \\
   &           & 153.0~(Var)  & 27 & 1707.2(0.3) &               \\
7  & 307.5(0.2)& 271(5)~(DMC) & 28 & 1813.4(0.4) &               \\
8  & 341.6(0.3)&              & 29 & 1825.8(0.4) &               \\
9  & 342.0(0.2)&              & 30 & 2446.8(0.3) & 2432(5)~(DMC) \\
10 & 342.9(0.2)&              & 31 & 2832.8(0.3) & 2974(5)~(DMC) \\
11 & 343.3(0.2)&              &    &             & 3023(5)~(DMC) \\
12 & 402.1(0.3)&              &    &             & 3144(5)~(DMC) \\
13 & 430.0(0.3)&              &    &             &               \\
\bottomrule
\end{tabular}
\end{table}

\subsection{On Levels Missing from the Comparison in Table~\ref{tab:full_energy}}
\label{subsec:missing_levels}

In the present NCT calculation, the same flow $f_\theta$ is applied to every function of the reference basis $\{\Phi_{\vec{n}}\}$ in \Eq{eq:Phi_z} to construct the variational wavefunctions, and the flow does not change the nodal number of a given basis function. Under this setup, eigenstates whose nodal number does not appear among the reference basis functions are expected to be less well captured by the NCT states in Table~\ref{tab:full_energy}.

On the DMC side, the fixed-node DMC of Ref.~\cite{hinkle_characterizing_2008} uses a manually designed nodal surface for each excited state, so specific states for which such a nodal surface is not available may likewise be missing from the DMC column of Table~\ref{tab:full_energy}.

\end{document}